\documentclass{jfm}

\usepackage{color}
\usepackage{graphicx}
\usepackage{newtxtext}
\usepackage{newtxmath}
\usepackage{natbib}
\usepackage{hyperref}
\usepackage{overpic}
\hypersetup{
    colorlinks = true,
    urlcolor   = blue,
    citecolor  = black,
}

\newcommand{\RomanNumeralCaps}[1]
\linenumbers

\newcommand{\vect}[1]{\boldsymbol{#1}}
\newcommand{\tens}[1]{\mathsfbi{#1}}

\title{A general model for spin coating on a non-axisymmetric curved substrate}

\author{Ross G. Shepherd\aff{1,2}\corresp{\email{rgs53@cam.ac.uk}}, Edouard Boujo\aff{3} \and Mathieu Sellier\aff{1} }

\affiliation{\aff{1}Department of Mechanical Engineering, University of Canterbury, Christchurch 8140, New Zealand
\aff{2}Institute for Energy and Environmental Flows, University of Cambridge, Cambridge CB3 0EZ, United Kingdom
\aff{3}Laboratory of Fluid Mechanics and Instabilities, \'Ecole Polytechnique F\'ed\'erale de Lausanne, Lausanne CH1015, Switzerland}

\begin{document}
\maketitle

\begin{abstract}
We derive a generalised asymptotic model for the flow of a thin fluid film over an arbitrarily-parameterised non-axisymmetric curved substrate surface based on the lubrication approximation. In addition to surface tension, gravity, and centrifugal force, our model incorporates the effects of the Coriolis force and disjoining pressure, together with a non-uniform initial condition, which have not been widely considered in existing literature. We use this model to investigate the impact of the Coriolis force and fingering instability on the spreading of a non-axisymmetric spin-coated film at a range of substrate angular velocities, first on a flat substrate, and then on parabolic cylinder- and saddle-shaped curved substrates. We show that, on flat substrates, the Coriolis force has a negligible impact at low angular velocities, and at high angular velocities results in a small deflection of fingers formed at the contact line against the direction of substrate rotation. On curved substrates, we demonstrate that as the angular velocity is increased, spin coated films transition from being dominated by gravitational drainage with no fingering to spreading and fingering in the direction with the greatest component of centrifugal force tangent to the substrate surface. For both curved substrates and all angular velocities considered, we show that the film thickness and total wetted substrate area remain similar over time to those on a flat substrate, with the key difference being the shape of the spreading droplet.
\end{abstract}

\begin{keywords}

\end{keywords}

\section{Introduction}
% Motivation
Spin coating is widely used to apply functional and protective coatings in the manufacturing of electronic and optical components, such as microprocessors, light-emitting diode displays, and solar panels.
The spin coating process consists of depositing a coating liquid onto a substrate surface, then rotating the substrate at high speed so that centrifugal force spreads the liquid over the surface \citep{Cohen2011}.
Once the liquid has formed a thin film over the entire substrate surface, the film is allowed to cure by solvent evaporation, photochemical, or other means, to leave a uniform and highly-reproducible coating.
Current spin coating techniques, however, are unable to reliably produce uniform coatings on curved substrates \citep{Rich2021}.
This restricts a wide range of spin-coated products to only flat geometries or singly-curved surfaces (by bending flat surfaces without stretching).

% Axisymmetric models
\citet{Emslie1958} developed a simple one-dimensional model for the dynamics of an axisymmetric thin fluid film on a rotating flat substrate, based on the lubrication approximation, considering only the effects of centrifugal force.
They showed that, regardless of the initial film profile, a spin coated film on a flat substrate will tend towards a uniform coating.
Models for spin-coated films have since been extended to incorporate additional effects such as gravitational and Coriolis forces, surface tension, and curved substrate geometry.
The case when surface-tension and moving-contact-line effects are significant on flat substrates was later studied by \citet{wilson2000rate}.
\citet{Chen2009} and \citet{Liu2017} presented one-dimensional models to predict the thickness of spin-coated films on convex spherical substrates, and demonstrated close agreement with the thickness of experimentally-measured films.
\citet{Kang2016} and \citet{Duruk2021} used similar models to investigate the flow of a film over the entire surface of rotating spherical and spheroidal substrates, paying particular attention to the film dynamics during the transition from gravity- to centrifugal force-driven flow as the angular velocity of the substrate is increased.
Finally, these one-dimensional models were generalised by \citet{Weidner2018} to allow for an arbitrary axisymmetric substrate, such as with ridges and dips.

% Non-axisymmetric films and fingering comparisons to experimetns
All of the above models have only considered the flow of axisymmetric films, and cannot capture the more complex dynamics which occur in non-axisymmetric flows, such as fingering instabilities, as observed experimentally by \citet{Fraysse1994}, and angular velocity components introduced by the Coriolis force.
In the case of an axisymmetric film, \citet{Myers2001} showed that the effects of the Coriolis force would have no impact on film thickness, but would induce an angular velocity component that could affect the evolution of non-axisymmetric films.
This was supported by experiments by \citet{Cho2005} demonstrating the deflection of fingers formed at the contact line due to the Coriolis force.
On a flat substrate, spreading from a droplet initial condition was investigated by \citet{Schwartz2004} without assuming an axisymmetric flow and including the Coriolis force, where they were able to reproduce the fingering instability observed by \citet{Fraysse1994} as well as finger deflection similar to \citet{Cho2005}, but only while using a viscosity 14 times smaller than used in experiments.

% Previous models of flows on arbitrary curved substrates
Better understanding the dynamics of thin liquid films on stationary curved substrates has been the topic of several studies starting with the work of \citet{schwartz1995modeling} who included the effects of substrate curvature in the lubrication approximation for planar flows. The effects of inertia were later considered in \citet{ruschak2003laminar} for a two-dimensional flow. For the three-dimensional flow of a thin liquid film on a stationary curved substrates, a general theory for thin films driven by surface tension and gravity was developed by \citet{Roy2002} and \citet{Thiffeault2006}, allowing for flow to be modelled on any smooth substrate geometry, and using any parameterisation of the substrate geometry. The possible inclusion of solidification in the governing equations was concurrently considered in \citet{myers2002flow} and the inertia effects later included in \citet{wray2017reduced}. This general framework was later used by \citet{Takagi2010}, \citet{Balestra2016,Balestra2018}, \citet{qin2021axisymmetric}, \citet{Ledda2022}, and \citet{mckinlay2023late} to study gravitational drainage and contact-line instabilities over a range of curved substrate geometries. The general theory was extended by \citet{Mayo2015} to simulate the dynamics of droplets on leaves (with the addition of disjoining pressure at the contact line).

Since the pioneering work of \citet{howell2003surface}, few studies have considered the combined effects of a non-trivial substrate kinematics with a complex substrate shape. A notable exception which builds on the large body of literature related to rimming flows on circular cylinder---see for example \citet{evans2004steady,rietz2017dynamics,lopes2018multiple,mitchell2022unsteady}---is the work in \citet{li2017viscous} investigating the free-surface dynamics of a thin film on a rotating elliptical cylinder. Recently, \citet{Duruk2023} modelled coating flows on rotating ellipsoids (with the addition of centrifugal force, but not the Coriolis force).

We aim, in this work, to extend these models by presenting an extended general theory which includes all non-inertial forces and therefore can reliably simulate spreading from a droplet initial condition over a rotating, non-axisymmetric curved substrate and shed light on the combined effects of rotation and substrate curvature on the film spreading dynamics.
We also aim to clarify to which extent the Coriolis force which has commonly been assumed (and convincingly been demonstrated) to be negligible for axisymmetric thin film flow configurations can still be ignored for non-axisymmetric surfaces. This has remained, to the best of the authors' knowledge, an open question.

% Structure of this paper
In section \ref{sec:model}, we will derive a dimensionless general lubrication model for the evolution of a thin fluid film over the surface of an arbitrarily-parameterised rotating curved substrate following a similar methodology to \cite{Roy2002} and \cite{Thiffeault2006}, incorporating the effects of surface tension, disjoining pressure, gravity, centrifugal, and Coriolis forces.

Section \ref{sec:implementation} gives the parameters and numerical details used to implement this model.
In section \ref{sec:results}, we present the results from a series of example simulations.
In section \ref{sec:flat substrate}, we first consider the spreading of a spin-coated droplet on a flat substrate in order to demonstrate the effects of the Coriolis force at different angular velocities.
In section \ref{sec:non-axisymmetric substrate}, we then show the effect of two different non-axisymmetric curved substrates (a parabolic cylinder and saddle) on the spreading of a spin-coated droplet and the onset of the fingering instability in gravity-driven, transitional, and centrifugal force-driven flow regimes.
Finally, in section \ref{sec:film thickness and coverage}, we present quantitative results comparing the rate of film spreading over the different substrate geometries.

\section{Model}\label{sec:model}
\subsection{Substrate-based curvilinear coordinate system}\label{sec:coordinate system}
% Tangent and Normal Vectors
Consider a thin film of an incompressible Newtonian fluid on a smooth substrate surface.
Let the substrate surface, $\vect{s}(x^1,x^2)$, be a two-dimensional Riemannian manifold parameterised by $(x^1,x^2)\in\mathbb{R}^2$.
Let
\begin{equation}
\vect{e}_\alpha =  \frac{\partial \vect{s}}{\partial x^\alpha}, \qquad
\hat{\vect{n}} = \frac{ \vect{e}_1 \times \vect{e}_2 }{\big\lVert \vect{e}_1 \times \vect{e}_2 \big\rVert},
\end{equation}
be basis vectors tangent to the substrate (for $\alpha \in \{1,2\}$, as with other Greek indices throughout), and a unit vector in the positive normal direction to the substrate, respectively.
Note that $\vect{e}_\alpha$ are not necessarily orthogonal or normalised, and a suitable choice of substrate parameterisation is required to ensure the desired surface orientation.
We also define cobasis vectors such that $\vect{e}_\alpha\vect{\cdot}\vect{e}^\beta = \delta_\alpha{}^\beta$ (where $\delta_\alpha{}^\beta$ is the Kronecker delta) and $\vect{e}^\alpha \vect{\cdot} \hat{\vect{n}} = 0$:
\begin{equation}
\vect{e}^1 = \frac{ \vect{e}_2 \times \hat{\vect{n}} }{\big\lVert \vect{e}_1 \times \vect{e}_2  \big\rVert}, \qquad
\vect{e}^2 = \frac{ \hat{\vect{n}} \times \vect{e}_1 }{\big\lVert \vect{e}_1 \times \vect{e}_2 \big\rVert}.
\end{equation}

\begin{figure}
\centering
\includegraphics[scale=0.9]{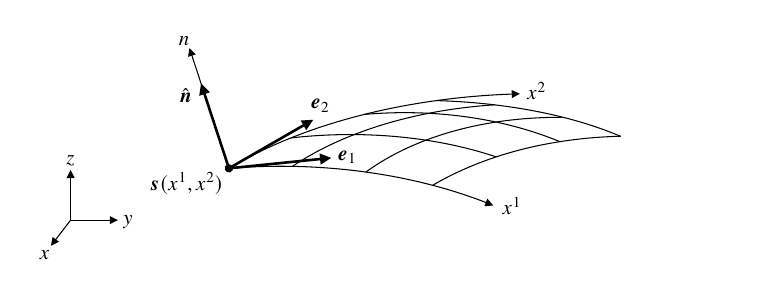}
\caption{The substrate surface and coordinate system, together with the basis vectors, $\vect{e}_1$, $\vect{e}_2$, and unit normal vector, $\hat{\vect{n}}$, at the point $\vect{s}(x^1,x^2)$.}
\label{fig:coord_system}
\end{figure}

% Define Normal Coordinate
Let $h(x^1,x^2,t)$ be the thickness of the fluid film above $\vect{s}(x^1,x^2)$ at time $t$, measured in the positive normal direction to the substrate surface.
Each point in the fluid film can then be written as
\begin{equation}
\vect{r}(x^1,x^2,n) = \vect{s}(x^1,x^2) + n\hat{\vect{n}}(x^1,x^2), \label{eq:position vector}
\end{equation}
where $0\leq n\leq h$ is the distance from the substrate surface.
This coordinate system is shown in figure \ref{fig:coord_system}.

% Metric and Curvature Tensors
Let $\tens{G}$ be the metric tensor on the substrate surface with components
\begin{equation}
G_{\alpha\beta} = \vect{e}_\alpha\vect{\cdot}\vect{e}_\beta,
\qquad G^{\alpha\beta} = \vect{e}^\alpha\vect{\cdot}\vect{e}^\beta.
\end{equation}
Distance and area elements on the substrate surface can then be written as
\begin{equation}
(\mathrm{d} s)^2 = G_{\alpha\beta}\,\mathrm{d} x^\alpha \mathrm{d} x^\beta, \qquad \mathrm{d} A = \sqrt{G}\,\mathrm{d} x^1 \mathrm{d} x^2,
\end{equation}
adopting the Einstein summation convention for repeated Greek indices, where $G = \det (\tens{G}) = G_{11} G_{22} - (G_{12})^2$ is the determinant of the metric.
Now, let
\begin{equation}\label{eq:curvature definition}
K_\alpha{}^\beta = \frac{\partial\vect{e}^\beta}{\partial x^\alpha}\vect{\cdot}\hat{\vect{n}}
\end{equation}
define the symmetric substrate curvature tensor, $\tens{K} = K_\alpha{}^\beta \vect{e}^\alpha \vect{e}_\beta$.
We can obtain the components $ K_{\alpha\beta} = G_{\beta\gamma} K_\alpha{}^\gamma$ and $K^{\alpha\beta} = G^{\gamma\beta} K_\gamma{}^\alpha$ by the usual way of raising and lowering indices with the metric.
We define the mean curvature and Gaussian curvature of the substrate, respectively, as follows:
\begin{equation}
\kappa =  \mathrm{tr}(\tens{K}) = K_\alpha{}^\alpha, \qquad K = \det (\tens{K}) = K_1{}^1 K_2{}^2 - K_1{}^2 K_2{}^1.
\end{equation}

% Extended Basis Vectors
Accounting for curvature, the basis and cobasis tangent vectors can be extended to the space around the substrate:
\begin{equation}
\vect{e}^+_\alpha = \frac{\partial \vect{r}}{\partial x^\alpha} = \vect{e}_\alpha - n K_\alpha{}^\beta \vect{e}_\beta, \qquad
\vect{e}^{+\alpha} = \vect{e}^\alpha + n K_\beta{}^\alpha \vect{e}^\beta + O(n^2),\label{eq:basis vector e+}
\end{equation}
so that $\vect{e}^+_\alpha\vect{\cdot}\vect{e}^{+\beta} = \delta_\alpha{}^\beta + O(n^2)$\footnote{Note that here $n^2$ refers to $n$ {\it squared} (and not the second contravariant component), since $n$ is scalar. This will be the case for powers of scalar quantities $h$, $L$, and $\varepsilon$, throughout, and should be apparent from context rather than use the notation $\{n\}^2$ or $(n)^2$.}.
This leads to an extended metric tensor parallel to the substrate surface, $\tens{G}^{\,+}$, with components
\begin{equation}\label{eq:metric G+}
\arraycolsep=1.5pt
\left. \begin{array}{c}
G^+_{\alpha\beta} = \vect{e}^+_\alpha\vect{\cdot}\vect{e}^+_\beta =  G_{\alpha\beta} - 2n K_{\alpha\beta} + O(n^2), \\[4pt]
G^{+\alpha\beta} = \vect{e}^{+\alpha}\vect{\cdot}\vect{e}^{+\beta} =  G^{\alpha\beta} + 2n K^{\alpha\beta} + O(n^2).
\end{array} \right\}
\end{equation}
An area element parallel to (but away from) the substrate surface is then
\begin{equation}
\sqrt{G^+} \, \mathrm{d} x^1 \mathrm{d} x^2 = \eta \sqrt{G} \, \mathrm{d} x^1 \mathrm{d} x^2 = \eta\,\mathrm{d} A,
\end{equation}
where the determinant of the extended metric is $G^+ = \det (\tens{G}^{\,+}) = \eta^2 G$, and the expansion or contraction of the coordinate system away from the substrate surface is characterised by
\begin{equation}
\eta = 1 - \kappa n + K n^2.
\end{equation}
Since the extended metric becomes non-invertible at $\eta=0$, the extended coordinate system is only valid when $\eta > 0$.
This condition is always satisfied on a convex or flat substrate surface, however when the substrate is concave in any direction (that is, when either of the eigenvalues, $k_1$, $k_2$, of $\tens{K}$ become strictly positive), this is only guaranteed when the normal coordinate is less than the minimum radius of curvature, $n < 1/\max(k_1,k_2)$.

\subsection{Governing equations and boundary conditions}
% NS Equations
Flow within the fluid film is governed by the Navier--Stokes equations (in coordinate-free form):
\begin{equation}
\vect{\nabla}\vect{\cdot}\vect{u} = 0,\label{eq:NS_conservation}
\end{equation}
\begin{equation}
\rho \bigg( \frac{\partial\vect{u}}{\partial t} + \vect{u}\vect{\cdot}\vect{\nabla}\vect{u} \bigg) = -\vect{\nabla} p +  \mu\Delta\vect{u} + \rho \vect{f},\label{eq:NS_momentum}
\end{equation}
where $\rho$ and $\mu$ are the density and viscosity of the fluid, $\vect{u}(x^1,x^2,n,t) = u^\alpha \vect{e}^+_\alpha + u^n\hat{\vect{n}}$ is the velocity in the fluid film (with contravariant components $u^\alpha$ tangent to the substrate, and component $u^n$ normal to the substrate), $p(x^1,x^2,n,t)$ is the pressure in the film, and $\vect{f}(x^1,x^2,n,t) = f^\alpha\vect{e}^+_\alpha + f^n\hat{\vect{n}}$ is the acceleration due to the total body force acting on the fluid (referred to herein as simply the body force, which may vary with both time and space depending on the substrate kinematics), and where $\vect{\nabla}$ and $\Delta = \vect{\nabla}\vect{\cdot}\vect{\nabla}$ are the usual three-dimensional gradient and Laplacian operators in $\mathbb{R}^3$.

% Continuity Equation
Let $\vect{q}(x^1,x^2,t) = q^\alpha \vect{e}_\alpha$ be the vector field of volumetric flux over the substrate surface with components
\begin{equation}
q^\alpha = \int_0^h \eta u^\alpha\,\mathrm{d} n. \label{eq:volume_flux_components}
\end{equation}
Integrating (\ref{eq:NS_conservation}) in $n$ with the boundary conditions $u^n\rvert_{n=0}=0$ and $u^n\rvert_{n=h} = \mathrm{d}h/\mathrm{d}t$ gives the following continuity equation in terms of the volume flux:
\begin{equation}
    \eta^* \frac{\mathrm{d} h}{\mathrm{d} t} + \vect{\nabla}_{\!S} \vect{\cdot} \vect{q} = 0, \label{eq:volume_balance}
\end{equation}
where $\eta^* = \eta\rvert_{n=h} = 1 - \kappa h + K h^2$ and $\vect{\nabla}_{\!S} = \vect{e}^\alpha \partial/\partial x^\alpha$ is the gradient operator over the substrate surface. The divergence over the substrate surface is given by \citep[see][p.\,78]{Lebedev2003}
\begin{equation}
\vect{\nabla}_{\!S} \vect{\cdot} \vect{q} = \vect{e}^\alpha \vect{\cdot} \frac{\partial \vect{q}}{\partial x^\alpha} = \frac{1}{\sqrt{G}}\frac{\partial}{\partial x^\alpha} \Big(\sqrt{G}\,q^\alpha\Big). \label{eq:substrate divergence}
\end{equation}

% Body Forces
For thin film flows in spin coating, we introduce the centrifugal and Coriolis forces induced by a substrate reference frame rotating at a constant speed \citep[p.\,461]{Morin2008}.
The total acceleration due to body forces acting on the fluid at the point $\vect{r}$ is then
\begin{equation}
\vect{f} = g \hat{\vect{g}} - \omega^2 \hat{\vect{\omega}} \times (\hat{\vect{\omega}} \times \vect{r}) - 2\omega\hat{\vect{\omega}} \times \vect{u},
\end{equation}
where $g$ is the acceleration due to gravity, $\hat{\vect{g}}$ is a unit vector in the direction of gravity, $\omega$ is the angular velocity of the substrate and $\hat{\vect{\omega}}$ is a unit vector in the direction of the axis of rotation (by the right-hand rule).

% Boundary Conditions
At the substrate surface, $n=0$, we impose the zero-slip boundary condition $\vect{u} = \vect{0}$.
At the free fluid surface, the velocity and pressure must satisfy the stress balance:
\begin{equation}\label{eq:stress balance}
\left. \begin{array}{c}
\mu \big( \hat{\vect{n}}^* \vect{\cdot} \tens{T} \vect{\cdot} \hat{\vect{n}}^* \big) = p - p_\text{a} + \gamma \kappa^* + \varPi, \\[4pt]
\mu \big( \vect{t}_\alpha^* \vect{\cdot} \tens{T} \vect{\cdot} \hat{\vect{n}}^* \big) = 0,
\end{array} \right\} \quad \text{on } n=h,
\end{equation}
where $\vect{t}_\alpha^*$ and $\hat{\vect{n}}^*$ are tangent and unit normal vectors to the free surface, $\tens{T} = \vect{\nabla}\vect{u} + \vect{\nabla}\vect{u}^\mathrm{T}$ is the strain rate tensor, $p_\text{a}$ is the ambient pressure, $\gamma$ is the surface tension at the fluid--air interface, $\kappa^*$ is the mean curvature of the free surface, and $\varPi$ is the disjoining pressure at the interface.
The disjoining pressure can be modelled as
\begin{equation}
\varPi = \frac{ \gamma (m_1-1)(m_2-1) }{ h_\text{p} (m_1-m_2) } ( 1 - \cos\theta_\text{e} ) \bigg[ \bigg(\frac{h_\text{p}}{h}\bigg)^{m_1} - \bigg(\frac{h_\text{p}}{h}\bigg)^{m_2} \bigg],
\end{equation}
where $h_\text{p}$ is a precursor film thickness, $\theta_\text{e}$ is the equilibrium contact angle between the fluid and the substrate, and $m_1$, $m_2$ are constants such that $m_1>m_2>1$.
In this paper we will use $m_1 = 3$ and $m_2 =2$ (as used by \citet{Mayo2015,Schwartz2001}).

\subsection{Non-dimensionalisation}\label{sec:non-dimesionalisation}
% Dimensionless Coordinate System, Basis Vectos, Metric, and Curvature
Let $h_\text{c}$ be the characteristic thickness of the fluid film, let $L$ be the characteristic length scale of the substrate, and let $\varepsilon=h_\text{c}/L \ll 1$ be the aspect ratio of the film.
We now define a dimensionless substrate coordinate system, rescaled by the characteristic lengths $h_\text{c}$ normal to the substrate and $L$ tangent to the substrate. This gives rise to rescaled position vectors, normal coordinate, and (co)basis vectors, temporarily indicated by a tilde:
\begin{equation}
\tilde{\vect{s}} = \frac{\vect{s}}{L}, \qquad
\tilde{\vect{r}} = \frac{\vect{r}}{L}, \qquad
\tilde{n} = \frac{1}{h_\text{c}}n, \qquad
\tilde{\vect{e}}_\alpha = \frac{1}{L}\vect{e}_\alpha, \qquad
\tilde{\vect{e}}^\alpha = L\vect{e}^\alpha, \label{eq:coordiante rescaling}
\end{equation}
and corresponding dimensionless metric and curvature tensors, $\tilde{\tens{G}}$ and $\tilde{\tens{K}}$, where
\begin{equation}
\tilde{G}_{\alpha\beta} = \frac{1}{L^2} G_{\alpha\beta}, \qquad
\tilde{G}^{\alpha\beta} = L^2  G^{\alpha\beta}, \qquad
\tilde{K}_\alpha{}^\beta = L K_\alpha{}^\beta.
\end{equation}
The dimensionless extended (co)basis vectors and metric away from the substrate surface can be written similarly to (\ref{eq:basis vector e+}) and (\ref{eq:metric G+}):
\begin{equation}
\arraycolsep=8pt
\left. \begin{array}{cc}
\tilde{\vect{e}}^+_\alpha = \tilde{\vect{e}}_\alpha - \varepsilon \tilde{n}\tilde{K}_\alpha{}^\beta \tilde{\vect{e}}_\beta, &
\tilde{\vect{e}}^{+\alpha} = \tilde{\vect{e}}^\alpha + \varepsilon \tilde{n}\tilde{ K}_\beta{}^\alpha \tilde{\vect{e}}^\beta + O(\varepsilon^2), \\[4pt]
\tilde{G}^+_{\alpha\beta} = \tilde{G}_{\alpha\beta} - 2\varepsilon\tilde{n}\tilde{ K}_{\alpha\beta} + O(\varepsilon^2),  &
\tilde{G}^{+\alpha\beta} = \tilde{G}^{\alpha\beta} + 2\varepsilon\tilde{n}\tilde{ K}^{\alpha\beta} + O(\varepsilon^2).
\end{array} \right\} \label{eq:basis vectors dimensionless}
\end{equation}
The determinants of the dimensionless metrics are
\begin{equation}
\tilde{G} = \det (\tilde{\tens{G}}) = \frac{1}{L^4} G, \qquad
\tilde{G}^+ = \det (\tilde{\tens{G}}^{\,+}) = \frac{1}{L^4} G^+ = \frac{\eta^2}{L^4} G,
\end{equation}
where $\eta$ can be expressed in terms of dimensionless quantities as
\begin{equation}
\eta = 1 - \varepsilon\tilde{\kappa}\tilde{n} + \varepsilon^2\tilde{K}\tilde{n}^2,
\end{equation}
with the dimensionless mean and Gaussian substrate curvatures:
\begin{equation}
\tilde{\kappa} = \mathrm{tr}(\tilde{\tens{K}}) = L\kappa, \qquad \tilde{K} = \det (\tilde{\tens{K}}) = L^2 K.
\end{equation}

% Dimensionless Variables
Let $f_\text{c}$ be the characteristic\footnote{ The choice of a suitable characteristic force scale depends on the substrate kinematics, as discussed in section \ref{sec:characteristic force scale}} acceleration due to body forces acting on the fluid, let $\rho L^2 h_\text{c}$ be a characteristic mass, and let $\rho L f_\text{c}$ be a characteristic pressure. Let $u_\text{c}$ and $t_\text{c}$ be characteristic velocity and time scales, as follows:
\begin{equation}\label{eq:characteristic u and t}
u_\text{c} = \frac{\rho h_\text{c}{}^2 f_\text{c}}{\mu}, \qquad t_\text{c} = \frac{L}{u_\text{c}} = \frac{\mu}{\varepsilon \rho h_\text{c} f_\text{c}}.
\end{equation}
This leads to the dimensionless variables:
\begin{equation}
\arraycolsep=8pt
\left. \begin{array}{cccc}
\displaystyle \tilde{h} = \frac{h}{h_\text{c}}, &
\displaystyle \tilde{t} = \frac{t}{t_\text{c}}, &
\displaystyle \tilde{p} = \frac{p - p_\text{a}}{\rho L f_\text{c}},\\[8pt]
\displaystyle \tilde{\vect{u}} = \frac{\vect{u}}{u_\text{c}}, &
\displaystyle \tilde{\vect{q}} = \frac{\vect{q}}{u_\text{c}h_\text{c}}, &
\displaystyle \tilde{\vect{f}} = \frac{\vect{f}}{f_\text{c}}.
\end{array} \right\} \label{eq:dimensionless variables}
\end{equation}
To be consistent with the definitions, $\tilde{\vect{u}} = \tilde{u}^\alpha\tilde{\vect{e}}^+_\alpha + \tilde{u}^n\hat{\vect{n}}$ and $\tilde{\vect{q}} = \tilde{q}^\alpha\tilde{\vect{e}}_\alpha$,  the components of the dimensionless velocity and flux are
\begin{equation}
\tilde{u}^\alpha = \frac{L u^\alpha}{u_\text{c}}, \qquad
\tilde{u}^n = \frac{u^n}{u_\text{c}}, \qquad
\tilde{q}^\alpha = \int_0^{\tilde{h}} \eta \tilde{u}^\alpha\,\mathrm{d} \tilde{n} = \frac{q^\alpha}{\varepsilon u_\text{c}}. \label{eq:dimensionless volume flux}
\end{equation}

\subsection{Dimensionless governing equations and boundary conditions}
% Dimensionless NS Momentum Equation
Substituting the dimensionless variables (\ref{eq:dimensionless variables}), the NS momentum equation (in coordinate-free form) can be re-written as
\begin{equation}
-\frac{1}{\rho f_\text{c}} \vect{\nabla} p + \frac{\mu}{\rho f_\text{c}} \Delta\vect{u} + \tilde{\vect{f}} = O(\varepsilon \mathit{Re}), \label{eq:NS momentum dimensionless}
\end{equation}
where $\mathit{Re} = \rho u_\text{c} h_\text{c} / \mu $ is the Reynolds number.
Furthermore, expanding and assuming that $\tilde{u}^n \sim \varepsilon \tilde{u}^\alpha$, we can express the Laplacian as \citep[see][p.\ 79]{Lebedev2003}
\begin{equation}\label{eq:velocity laplacian}
\Delta\vect{u}
= \frac{\rho f_\text{c}}{\mu}
\bigg( \frac{\partial^2 \tilde{u}^\alpha}{\partial \tilde{n}^2}
- \varepsilon \big( \tilde{\kappa} \delta_\beta{}^\alpha + 2 \tilde{K}_\beta{}^\alpha \big) \frac{\partial \tilde{u}^\beta}{\partial \tilde{n}} \bigg)\tilde{\vect{e}}^+_\alpha
+ O(\varepsilon)\hat{\vect{n}} + O(\varepsilon^2)\tilde{\vect{e}}^+_\alpha,
\end{equation}
and the pressure gradient as \citep[see][p.\ 63]{Lebedev2003}
\begin{equation}\label{eq:pressure gradient}
\vect{\nabla}p = \rho f_\text{c} \bigg(
\tilde{\vect{e}}^{+\alpha}\frac{\partial\tilde{p}}{\partial x^\alpha} + \hat{\vect{n}}\frac{1}{\varepsilon}\frac{\partial \tilde{p}}{\partial\tilde{n}}
\bigg).
\end{equation}
Substituting these into (\ref{eq:NS momentum dimensionless}), the normal and tangential components of the NS momentum equation can be simplified to
\begin{equation}\label{eq:NS lubrication}
\left. \begin{array}{c}
\displaystyle{
-\frac{\partial \tilde{p}}{\partial \tilde{n}} + \varepsilon\tilde{f}^n = O(\varepsilon^2), }\\[8pt]
\displaystyle{
-\tilde{ G}^{+\alpha\beta}\frac{\partial\tilde{p}}{\partial x^\beta} + \frac{\partial^2\tilde{u}^\alpha}{\partial\tilde{n}^2}
- \varepsilon \big( \tilde{\kappa} \delta_\beta{}^\alpha + 2 \tilde{K}_\beta{}^\alpha \big) \frac{\partial \tilde{u}^\beta}{\partial \tilde{n} \vphantom{n^2}}
+ \tilde{f}^\alpha = O(\varepsilon\mathit{Re},\varepsilon^2). }
\end{array} \right\}
\end{equation}

% Dimensionless Continuity Equation
Substituting the dimensionless variables (\ref{eq:dimensionless variables}) into (\ref{eq:volume_balance}) gives the dimensionless continuity equation:
\begin{equation}\label{eq:volume balance dimensionless}
0 = \eta^* \frac{\partial\tilde{h}}{\partial\tilde{t}} + \tilde{\vect{\nabla}}_{\!S} \vect{\cdot}\tilde{\vect{q}},
\end{equation}
where $\eta^* = 1 - \varepsilon\tilde{\kappa}\tilde{h} + \varepsilon^2\tilde{K}\tilde{h}^2$ and $\tilde{\vect{\nabla}}_{\!S} = L \vect{\nabla}_{\!S}$ gives the dimensionless divergence over the substrate surface:
\begin{equation}\label{eq:divergence dimensionless}
\tilde{\vect{\nabla}}_{\!S} \vect{\cdot} \tilde{\vect{q}} = \tilde{\vect{e}}^\alpha \vect{\cdot} \frac{\partial \tilde{\vect{q}}}{\partial x^\alpha} = \frac{1}{\sqrt{\tilde{G}}}\frac{\partial}{\partial x^\alpha} \Big(\sqrt{\tilde{G}}\,\tilde{q}^\alpha\Big).
\end{equation}

% Dimensionless Body Forces
Rescaled against the characteristic body force, $f_\text{c}$, and substituting the dimensionless position vector, $\tilde{\vect{r}} = \tilde{\vect{s}} + \varepsilon \tilde{n} \hat{\vect{n}}$, the dimensionless body force is
\begin{equation}
\tilde{\vect{f}} = N_\text{grav} \hat{\vect{g}}
- N_\text{cent} \hat{\vect{\omega}} \times (\hat{\vect{\omega}} \times \tilde{\vect{s}})
- \varepsilon\tilde{n} N_\text{cent} \hat{\vect{\omega}} \times (\hat{\vect{\omega}} \times \hat{\vect{n}})
- 2 \varepsilon \mathit{Ta} \,\hat{\vect{\omega}} \times \tilde{\vect{u}},
\end{equation}
where $N_\text{grav} = g/f_\text{c}$ and $N_\text{cent} = \omega^2 L/f_\text{c}$ are dimensionless groups describing the ratio of gravity and centrifugal force to the characteristic force scale, and $\mathit{Ta} = \omega \rho L h_\text{c} / \mu $ is the Taylor number, describing the ratio of angular momentum to viscous forces and characterising the strength of the Coriolis force.

% Dimensionless Boundary Conditions
In dimensionless form, the zero-slip condition is $\tilde{\vect{u}} = \vect{0}$ on $\tilde{n}=0$, and the stress balance simplifies to
\begin{equation}\label{eq:dimensionless no shear boundary unsimplified}
\left. \begin{array}{c}
\tilde{p} = -N_\text{surf} \tilde{\kappa}^* - \tilde{\varPi} + O(\varepsilon^2) ,\\[4pt]
\displaystyle\frac{\partial\tilde{u}^\alpha}{\partial \tilde{n}} = O(\varepsilon^2),
\end{array} \right\} \quad \text{on }\tilde{n}=\tilde{h},
\end{equation}
where $N_\text{surf} = \gamma / \rho L^2 f_\text{c}$ is a dimensionless group describing the ratio of surface tension to the characteristic force scale, and $\tilde{\varPi} = \varPi/\rho L f_\text{c}$ is the dimensionless disjoining pressure:
\begin{equation}
\tilde{\varPi} = \frac{ 2 N_\text{surf} }{\varepsilon \tilde{h}_\text{p}} ( 1 - \cos\theta_\text{e} ) \bigg[ \bigg(\frac{\tilde{h}_\text{p}}{\tilde{h}}\bigg)^3 - \bigg(\frac{\tilde{h}_\text{p}}{\tilde{h}}\bigg)^2 \bigg],
\end{equation}
where $\tilde{h}_\text{p} = h_\text{p}/h_\text{c}$ is the dimensionless precursor film thickness.
$N_\text{surf}$ can take the form of different common dimensionless groups depending on the choice of force scale, as discussed in section \ref{sec:characteristic force scale}.
The dimensionless free surface curvature can be approximated as
\begin{equation}
\tilde{\kappa}^* = \tilde{\kappa} + \varepsilon \tilde{\kappa}_2 \tilde{h} + \varepsilon \tilde{\Delta}_{S} \tilde{h} + O(\varepsilon^2), \label{eq:free surface curvature}
\end{equation}
where $\tilde{\kappa}_2 = \tilde{K}_\alpha{}^\beta \tilde{K}_\beta{}^\alpha$, and $\tilde{\Delta}_{S} = \tilde{\vect{\nabla}}_{\!S} \cdot \tilde{\vect{\nabla}}_{\!S}$ is the dimensionless Laplacian over the substrate surface.
In the absence of disjoining pressure, these boundary conditions are equivalent to those used by \citet{Roy2002} and \citet{Thiffeault2006}.

\subsection{Choice of characteristic force scale}\label{sec:characteristic force scale}
The form of the dimensionless groups, $N_\text{surf}$, $N_\text{grav}$, and $N_\text{cent}$, characterising the strength of surface tension, gravitational, and centrifugal forces, is determined by the choice of the characteristic acceleration due to body forces acting on the fluid film, $f_\text{c}$.
Depending on the fluid properties, film thickness, substrate geometry, and kinematics, the dominant force acting on the film may be any of surface tension, gravity, or centrifugal force.
Each of these could therefore be justifiably chosen as a characteristic force scale when modelling different applications.
In each case, one of $N_\text{surf}$, $N_\text{grav}$, and $N_\text{cent}$ reduces to 1, and the others to well-known dimensionless groups---the Bond number, rotational Weber number, and rotational Froude number or their reciprocals---as summarised in table \ref{tab:dimensionless groups}.

\begin{table}
\centering
\includegraphics[scale=0.9]{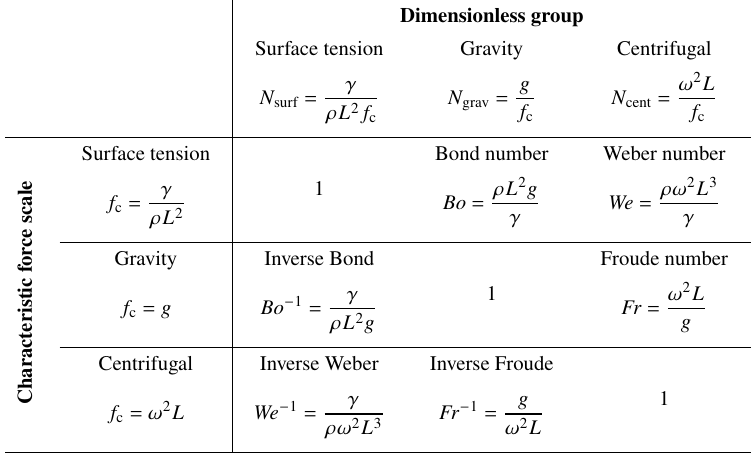}
\caption{Dimensionless groups resulting from different choices of characteristic force scale.}\label{tab:dimensionless groups}
\end{table}

In existing literature, surface tension has typically been chosen as a force scale when modelling very thin films and droplets, including \citet{Mayo2015}, \citet{Roy2002}, and \citet{Thiffeault2006}.
Gravity has been chosen as a force scale when considering the drainage of films over curved substrates, such as in \citet{Balestra2016,Balestra2018}, \citet{Ledda2022}, and \citet{Takagi2010}.
Finally, centrifugal force is a natural choice of force scale when modelling spin coating at high speeds, as used in \citet{Emslie1958} and \citet{Liu2017}.
When considering situations where several of these forces have a comparable effect on the film dynamics, there is not always a clear choice of force scale.
In this case, we propose a more general choice of characteristic force:
\begin{equation} \label{eq:general characteristic force}
f_\text{c} = \frac{\gamma}{\rho L^2} + g + \omega^2 L,
\end{equation}
which ensures that $N_\text{surf} + N_\text{grav} + N_\text{cent} = 1$.
This allows the dimensionless groups to be interpreted as the relative strength of each force.
In the limiting cases of flow driven entirely by surface tension, gravity, or centrifugal force, (\ref{eq:general characteristic force}) reduces to one of the characteristic force scales listed in table \ref{tab:dimensionless groups}.
Using this generalised force scale, we can smoothly transition between appropriate scalings for gravity- and centrifugal force-driven flows in order to reconcile the differing timescales of the regimes demonstrated in section \ref{sec:results}.

\subsection{General lubrication model}
Integrating the NS equations using a perturbation expansion approach similar to \citet{Roy2002} and \citet{Thiffeault2006} (see appendix \ref{app:integration of NS equations}), we obtain an expression for the components of the dimensionless volume flux over the substrate surface (omitting the tildes denoting dimensionless variables):
\begin{equation}\label{eq:general volume flux components}
\begin{split}
q^\alpha
= \frac{h^3}{3} \bigg[ & \bigg( \delta_\beta{}^\alpha - \varepsilon h \bigg( \kappa \delta_\beta{}^\alpha - \frac{1}{2} K_\beta{}^\alpha \bigg) - \varepsilon h^2 \omega^n \frac{4 \mathit{Ta}}{5} \epsilon_\beta{}^\alpha \bigg) \nabla^\beta (N_\text{surf}\kappa^* + \varPi) \\
& + \bigg( \delta_\beta{}^\alpha - \varepsilon h \bigg( \kappa \delta_\beta{}^\alpha + \frac{1}{2} K_\beta{}^\alpha \bigg) - \varepsilon h^2 \omega^n \frac{4 \mathit{Ta}}{5} \epsilon_\beta{}^\alpha \bigg) f^\beta_{(0)} \\
& + \varepsilon f^n_{(0)} \nabla^\alpha h 
+ \varepsilon h N_\text{cent} ( \hat{\vect{\omega}} \times ( \hat{\vect{\omega}} \times \hat{\vect{n}} )) \vect{\cdot} \vect{e}^\alpha
\bigg]
+ O(\varepsilon\mathit{Re}, \varepsilon^2),
\end{split}
\end{equation}
where $\nabla^\alpha = G^{\alpha\beta} \partial/\partial x^\beta$ are the contravariant components of the substrate gradient, $f^\alpha_{(0)}$ and $f^n_{(0)}$ are the leading-order components of the total body force tangent and normal to the substrate surface:
\begin{equation}
f^\alpha_{(0)} = \big[
N_\text{grav} \hat{\vect{g}}
- N_\text{cent} \hat{\vect{\omega}} \times ( \hat{\vect{\omega}} \times \vect{s} )
\big] \vect{\cdot} \vect{e}^{\alpha},
\end{equation}
\begin{equation}
f^n_{(0)} = \big[
N_\text{grav} \hat{\vect{g}}
- N_\text{cent} \hat{\vect{\omega}} \times ( \hat{\vect{\omega}} \times \vect{s} )
\big] \vect{\cdot} \hat{\vect{n}},
\end{equation}
and $\epsilon_\beta{}^\alpha$ is a mixed component of the modified Levi--Civita tensor, defined by
\begin{equation}\label{eq:levi-civita symbol}
\epsilon_{\alpha\beta}
= ( \vect{e}_\alpha \times \vect{e}_\beta ) \vect{\cdot} \hat{\vect{n}}
= \big\lVert \vect{e}_1 \times \vect{e}_2 \big\rVert
\begin{cases}
1 & \text{if }\alpha=1,\,\beta=2, \\
-1 & \text{if }\alpha = 2,\,\beta=1, \\
0 & \text{otherwise.}
\end{cases}
\end{equation}
Together with (\ref{eq:volume balance dimensionless}), this gives a partial differential equation (PDE) describing the evolution of the thickness, $h$, of a thin fluid film on a arbitrary rotating curved substrate.
In the absence of substrate rotation, (\ref{eq:general volume flux components}) is equivalent to equation (3.13) from \citet{Mayo2015}.
Furthermore, in the absence of disjoining pressure, (\ref{eq:general volume flux components}) is equivalent to equation (51) from \citet{Roy2002} and equations (IV.12)-(IV.14) from \citet{Thiffeault2006}.

\section{Numerical implementation}\label{sec:implementation}
\subsection{Simulation parameters}
For the example simulations in section \ref{sec:results}, we choose physical parameters and material properties based on spin coating experiments using silicon oil by \citet{Wang2001} and \citet{Cho2005}, as shown in table \ref{tab:physical parameters}.
The characteristic film thickness was chosen as $h_\text{c} = V/L^2$, so that the initial droplet has a dimensionless volume of $1$.
For a range of substrate angular velocities from $\omega=0\,\text{rad/s}$ to $200\,\text{rad/s}$, the corresponding dimensionless groups and characteristic time scales are shown in table \ref{tab:dimensionless groups varied omega}.
With this choice of parameters, we see from the relative magnitudes of $N_\text{surf}$, $N_\text{grav}$, and $N_\text{cent}$ that the dynamics will transition from gravity-dominated to centrifugal force-dominated over the range of $\omega$ considered here, with surface tension always having a small effect.

\begin{table}
\centering
\begin{tabular}{lcc}
Parameter & Symbol & Value \\
\hline
Total volume of initial droplet         & $V$		        & $0.5$\,\text{mL}\\
Characteristic substrate length         & $L$ 				& $50\,\text{mm}$ \\
Characteristic film thickness           & $h_\text{c}$ 		& $0.2\,\text{mm}$ \\
Dimensionless initial droplet height    & $h_0$             & $6.72$ \\
Dimensionless initial droplet radius of curvature     & $r_0$             & $1.77$ \\
Dimensionless precursor film thickness  & $h_\text{p}$      & $0.1$ \\
Film aspect ratio                       & $\varepsilon$     & $4\times 10^{-3}$ \\
Density                                 & $\rho$			& $980\,\text{kg/m}^3$ \\
Viscosity                               & $\mu$             & $1\,\text{Pa\,s}$ \\
Fluid--air surface tension              & $\gamma$			& $18.4\,\text{mN/m}$ \\
Equilibrium contact angle               & $\theta_\text{e}$	& $10^\circ$ \\
Gravitational acceleration              & $g$				& $9.81\,\text{m/s}^2$
\end{tabular}
\caption{Physical parameters and fluid properties used throughout section \ref{sec:results} (based on \citet{Wang2001} and \citet{Cho2005}).}\label{tab:physical parameters}
\end{table}

\begin{table}
\centering
\renewcommand{\arraystretch}{1.25}
\begin{tabular}{ccccccc}
$\omega$ & $N_\text{surf}$ & $N_\text{grav}$ & $N_\text{cent}$ & $\mathit{Re}$ & $\mathit{Ta}$ &  $t_\text{c}$ \\
\hline
0 rad/s		& $7.6 \times 10^{-4}$	& 1.00	& 0		& $7.5\times 10^{-5}$	& 0		& $129.9\,\text{s}$ \\
25 rad/s    & $1.8 \times 10^{-4}$	& 0.24	& 0.76	& $3.2\times 10^{-4}$	& 0.25 	& $31.1\,\text{s}$ \\
50 rad/s	& $5.6 \times 10^{-5}$	& 0.07	& 0.93	& $1.0\times 10^{-3}$	& 0.49 	& $9.5\,\text{s}$ \\
100 rad/s	& $1.5 \times 10^{-5}$	& 0.02	& 0.98	& $3.9\times 10^{-3}$	& 0.98	& $2.5\,\text{s}$ \\
200 rad/s   & $3.7 \times 10^{-6}$  & 0.005 & 0.99  & 0.015                 & 1.96  & $0.63\,\text{s}$ \\
\end{tabular}
\caption{Dimensionless groups and characteristic time scale corresponding to different angular velocities with all other parameters as listed in table \ref{tab:physical parameters}.}\label{tab:dimensionless groups varied omega}
\end{table}

\subsection{Initial conditions}
\begin{figure}
\centering
\includegraphics[scale=0.9]{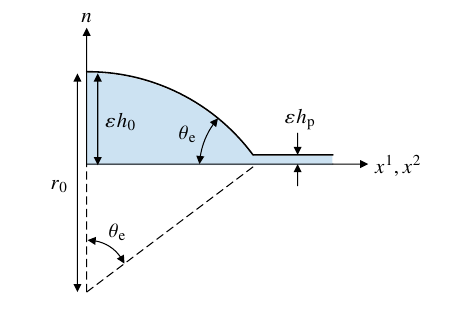}
\caption{Radial cross-section of a spherical cap initial condition (scaled by $L$) with dimensionless radius of curvature $r_0$, maximum height $h_0$, and precursor film of thickness $h_\text{p}$.}
\label{fig:initial_condition}
\end{figure}

For the example simulations, we choose a spherical cap initial condition with dimensionless radius of curvature $r_0$, maximum height $h_0$, and surrounding precursor film height $h_\text{p}$, as shown in figure \ref{fig:initial_condition}.
The (dimensional) volume of the spherical cap is given by \citep[p.~69]{Polyanin2007}
\begin{equation}
    V = \frac{\pi h_\text{c}{}^3 h_0{}^3}{3} (3 L r_0 - h_\text{c} h_0). \label{eq:cap volume dimensional r h}
\end{equation}
Expressing the maximum height in terms of the contact angle, $h_\text{c} h_0 = L r_0 (1-\cos\theta_\text{e})$, and setting $V = h_\text{c}L^2$ to ensure a dimensionless volume of $1$ (excluding the precursor film), (\ref{eq:cap volume dimensional r h}) can be re-written as:
\begin{equation}
    \frac{\pi L^3 r_0{}^3}{3}(2+\cos\theta_\text{e})(1-\cos\theta_\text{e})^2 = h_\text{c}L^2.
\end{equation}
Finally, rearranging and recalling that $\varepsilon = h_\text{c}/L$ is the film aspect ratio, the radius of curvature and maximum height of a spherical cap initial condition must satisfy
\begin{equation}
    {\frac{\pi r_0{}^3}{3\varepsilon}(2+\cos\theta_\text{e})(1-\cos\theta_\text{e})^2} = 1, \qquad
    h_0 = \frac{r_0}{\varepsilon}(1-\cos \theta_\text{e}).
\end{equation}
For a contact angle of $\theta_\text{e} = 10^\circ$, this results in $r_0 = 1.77$ and $h_0 = 6.72$.

We choose a dimensionless precursor film thickness of $h_\text{p} = 0.1$ (equivalent to $1.5\%$ of the initial droplet height) as a practical consideration for efficient simulations (see appendix~\ref{app:precursor} for a discussion of the effect of the precursor film thickness on the rate of film spreading, and also note that \citet{Spaid1996} showed that the choice of precursor film thickness does not affect the most amplified wavelength in fingering at the contact line).
For all of the examples in section \ref{sec:results}, we will always parameterise the substrate surface with $(x,y) = (x^1,x^2)$.
In this case, the initial film thickness for a spherical cap with a precursor film is given by
\begin{equation}\label{eq:h(0) initial condition}
    h(x^1,x^2,0) = \max\bigg\{\sqrt{r_0{}^2-(x^1)^2-(x^2)^2} - r_0 + h_0, h_\text{p} \bigg\}.
\end{equation}

We also consider the flow of a droplet spreading from a randomly perturbed initial condition.
To introduce perturbations with a range of wavelengths, we replace $r_0$ in (\ref{eq:h(0) initial condition}) with
\begin{equation}\label{eq:perturbation initial condition}
    r(\theta) = r_0\bigg( 1 + \sum_{n=1}^{50} a_n \sin (n\theta) + b_n \cos(n\theta) \bigg),
\end{equation}
where $\theta$ is the polar angle in the $x^1x^2$-plane and $a_n, b_n$ are normally-distributed random coefficients with a mean of $0$ and standard deviation of $5\times 10^{-3}$.

\subsection{Implementation using COMSOL Multiphysics and MATLAB}
Throughout section \ref{sec:results}, equations (\ref{eq:volume balance dimensionless}) and (\ref{eq:general volume flux components}) are solved using the finite-element {\it Coefficient Form PDE Interface} in COMSOL Multiphysics 5.6 with LiveLink for MATLAB R2021b.
In order to implement (\ref{eq:volume balance dimensionless}) and (\ref{eq:general volume flux components}), we introduce the variable $\varGamma = N_\text{surf}\kappa^* + \varPi$ and express the problem as a system of second-order PDEs in $h$ and $\varGamma$.
The system of PDEs is then solved by pre-computing the coefficients (including the components of the total body force, curvature tensor, and modified Levi--Civita tensor) over the solution mesh in MATLAB, then using COMSOL Multiphysics with all default settings and a solver tolerance of $10^{-5}$ (to ensure that the tolerance is less than $\varepsilon^2$ for the chosen value of $\varepsilon = 4\times 10^{-3}$) to solve for the evolution of $h$ and $\varGamma$ over time.
For the example simulations, we use a $200 \times 200$ cell square mesh with linear Lagrange elements over a domain of $(x^1,x^2) \in [-1,1] \times [-1,1]$.
Additionally, we choose zero-flux conditions on the boundaries of the computational domain.
This does not affect the spreading of the initial droplet as we do not allow any simulations to run sufficiently long for the droplet to reach the edge of the domain.

\section{Results}\label{sec:results}
\subsection{Effects of the Coriolis force in spin coating on a flat substrate}\label{sec:flat substrate}
\begin{figure}
    \centering
    \includegraphics{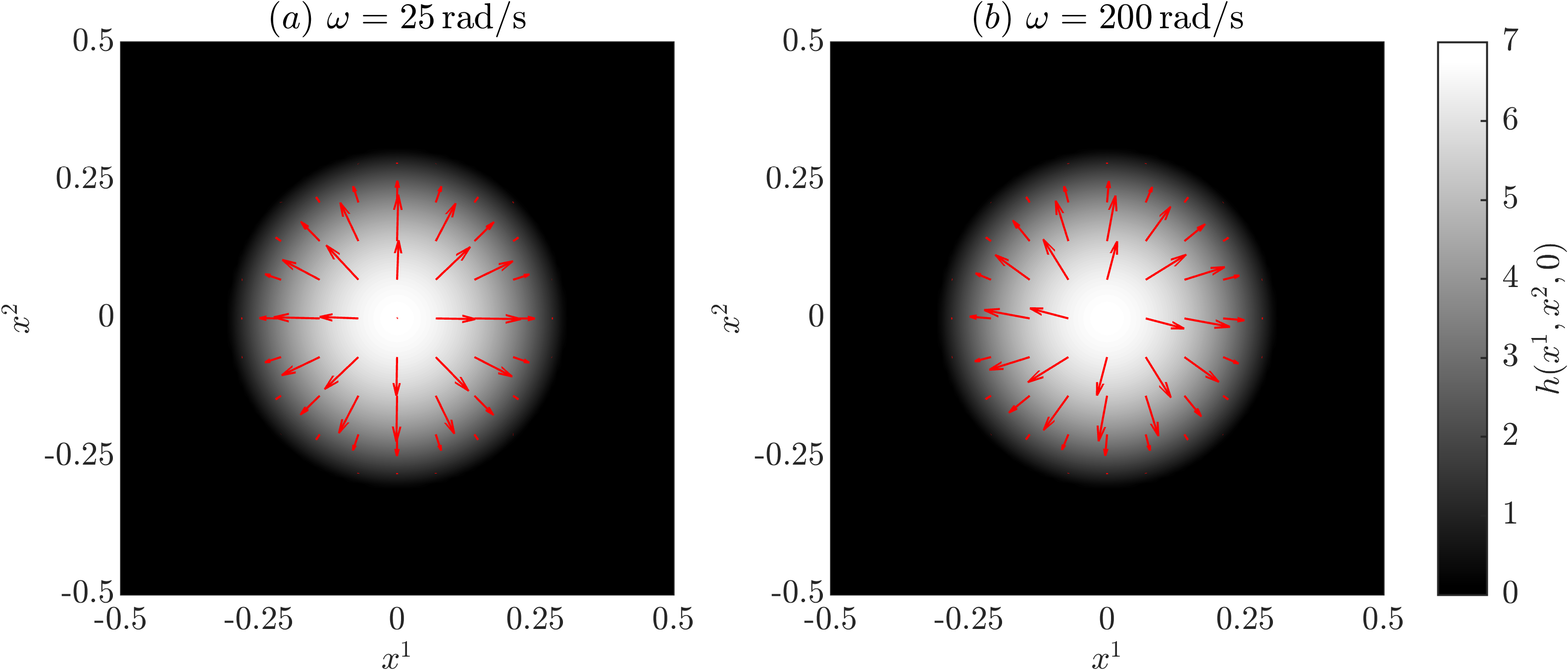}
    \caption{Vector field of the initial dimensionless volume flux, $\vect{q}(x^1,x^2,0)$, and colourmap of initial dimensionless film thickness, $h(x^1,x^2,0)$, for (\textit{a}) transitional flow ($\omega = 25\,\text{rad/s}$, $\mathit{Ta} = 0.25$) and (\textit{b}) centrifugal force-driven flow ($\omega = 200\,\text{rad/s}$, $\mathit{Ta} = 1.96$) on an anticlockwise-rotating substrate.}
    \label{fig:Initial flux vector field flat}
\end{figure}

\begin{figure}
    \centering
    \includegraphics{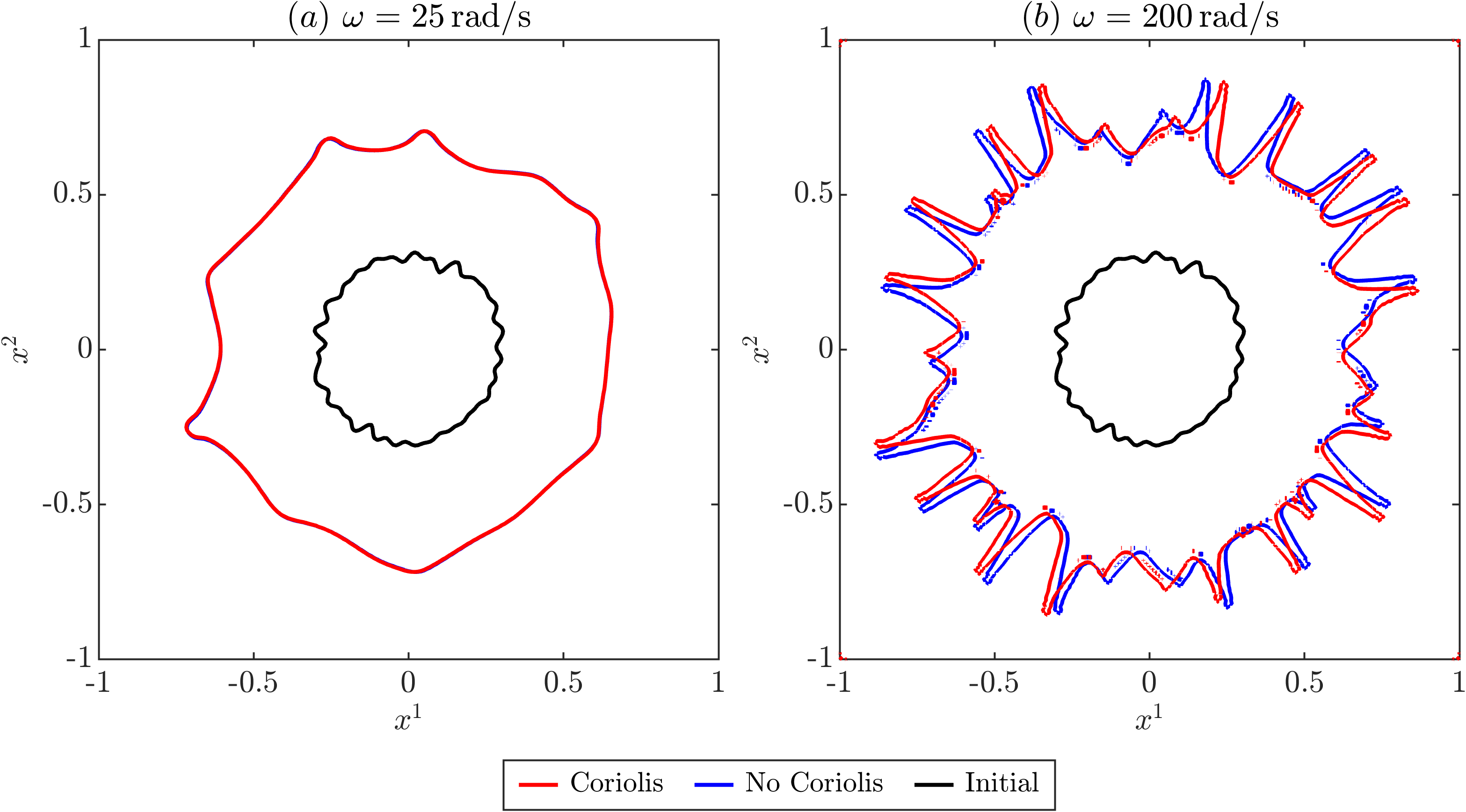}
    \caption{Contact line at $t=1.5$ with (red) and without (blue) the effects of the Coriolis force on a flat substrate from a randomly perturbed initial condition (black) on an anticlockwise-rotating substrate.
    (\textit{a}) Transitional flow ($\omega = 25\,\text{rad/s}$, $\mathit{Ta} = 0.25$), showing the near-indistinguishable contact lines with and without the Coriolis force.
    (\textit{b}) Centrifugal force-driven flow ($\omega = 200\,\text{rad/s}$, $\mathit{Ta} = 1.96$), demonstrating the deflection of radial fingers against the direction of substrate rotation due to the onset of the Coriolis force.}
    \label{fig:Final contact line flat perturbation}
\end{figure}

Before investigating the flow of spin-coated films on curved substrates, we will fist consider the base case of flow on a flat substrate with particular attention to the effect of the Coriolis force for the physical parameters considered here, which has not been included in previous models for flows on curved substrates (e.g.~\citet{Weidner2018,Duruk2023}).
Figure \ref{fig:Initial flux vector field flat} shows the instantaneous initial volume flux in a spherical droplet (\ref{eq:h(0) initial condition}) on a rotating flat substrate (given in dimensionless Cartesian coordinates by $\vect{s}(x^1,x^2) = [x^1,x^2,0]$) in two different flow regimes: transitional flow at low angular velocity ($\omega = 25\,\text{rad/s}$, $\mathit{Ta}=0.25$), where gravitational and centrifugal forces have a comparable effect; and centrifugal force-driven flow at high angular velocity ($\omega = 200\,\text{rad/s}$, $\mathit{Ta}=1.96$).
In the transitional regime (figure \ref{fig:Initial flux vector field flat}\textit{a}), the volume flux is entirely in the radial direction.
In the centrifugal force-driven regime (figure \ref{fig:Initial flux vector field flat}\textit{b}), however, the volume flux is deflected in the clockwise direction due to the onset of the Coriolis force, against the anti-clockwise direction of substrate rotation.
Compared to gravity and the leading-order component of centrifugal force, which do not scale with film thickness, the Coriolis force scales with $h^2$ (as seen in (\ref{eq:general volume flux components}) and detailed in the derivation in appendix~\ref{app:integration of NS equations}).
The effect will therefore be the strongest during the earliest stages of the spin coating process (as seen in the instantaneous initial volume flux), and diminish as the droplet spreads and thins.
Figure \ref{fig:Final contact line flat perturbation} shows the contact line at $t=1.5$ with and without the effect of the Coriolis force (by setting $\mathit{Ta}=0$) for the spreading of a perturbed spherical droplet (\ref{eq:perturbation initial condition}) on a flat substrate in transitional and centrifugal force-driven regimes.
Here, we plot the contact line as the contour where $h = 5h_\text{p} = 0.5$, noting that the dimensionless film thickness in the wetted area remains above 0.8 even at the end of our simulations.
Again, we see that at low angular velocity there is no appreciable difference in the contact line due to the Coriolis force.
With increasing angular velocity, the introduction of the Coriolis force leads to a slight deflection of the fingering at the contact line against the direction of substrate rotation with no change in the wavelength or amplitude of the fingering instability.

In the case of centrifugal-force dominated flow, where $f_\text{c} \approx \omega^2 L$ (and recalling the choice of characteristic velocity from (\ref{eq:characteristic u and t})), the Reynolds number may be expressed as
\begin{equation}
    \mathit{Re} = \frac{ \rho u_\text{c} h_\text{c} }{\mu}
    \approx \varepsilon \bigg(\frac{\omega \rho L h_\text{c}}{\mu}\bigg)^2
    = \varepsilon \mathit{Ta}^2.
\end{equation}
This sets an upper bound of $\mathit{Ta} \lesssim 1$ in order to ensure that $\mathit{Re} \lesssim \varepsilon$, which limits the higher-order terms in (\ref{eq:general volume flux components}) to at most $O(\varepsilon^2)$.
There is therefore a narrow range of angular velocities around $\mathit{Ta}\sim 1$ where the Coriolis force has an observable effect on the flow (such as in figures \ref{fig:Initial flux vector field flat}\textit{b} and \ref{fig:Final contact line flat perturbation}\textit{b}), but where inertial effects can still be ignored.
For the model and parameters considered here, we can conclude that while it is important to include the Coriolis force in order to accurately model the film dynamics, its effect is small enough that it is not likely to have a significant impact in practical applications.

\subsection{Spin coating on non-axisymmetric curved substrates}\label{sec:non-axisymmetric substrate}
We now consider the flow of a spin-coated film on two different non-axisymmetric curved substrates.
In sections \ref{sec:parabolic cylinder results} and \ref{sec:saddle results}, we will discuss example simulations of a perturbed spherical droplet spreading on rotating parabolic cylinder- and saddle-shaped substrates, respectively.
In each case, the results from additional realisations of the randomly-perturbed initial condition are reported in appendix~\ref{app:random}.

\begin{figure}
    \centering
    \includegraphics{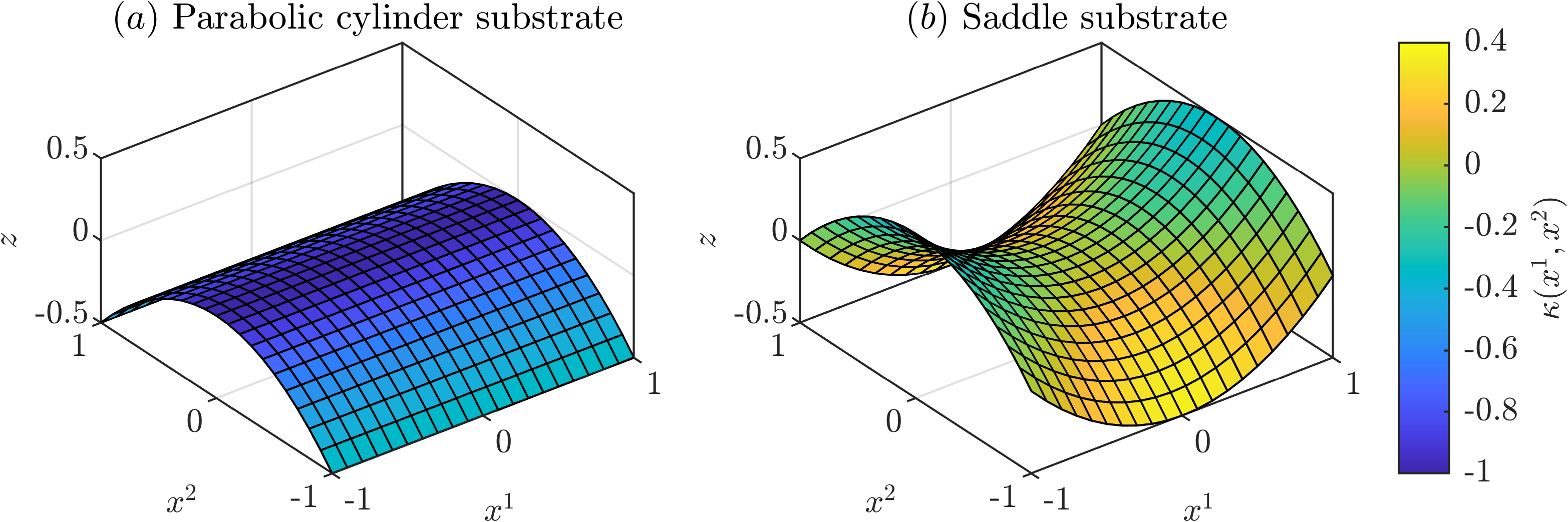}
    \caption{(\textit{a}) Parabolic cylinder substrate (\ref{eq:substrate geometry cylinder}) and (\textit{b}) saddle substrate (\ref{eq:substrate geometry saddle}) in dimensionless Cartesian coordinates, coloured by the dimensionless substrate mean curvature, $\kappa(x^1,x^2)$.}
    \label{fig:Substrate geometry}
\end{figure}

\subsubsection{Parabolic cylinder substrate}\label{sec:parabolic cylinder results}
\begin{figure}
    \centering
    \includegraphics{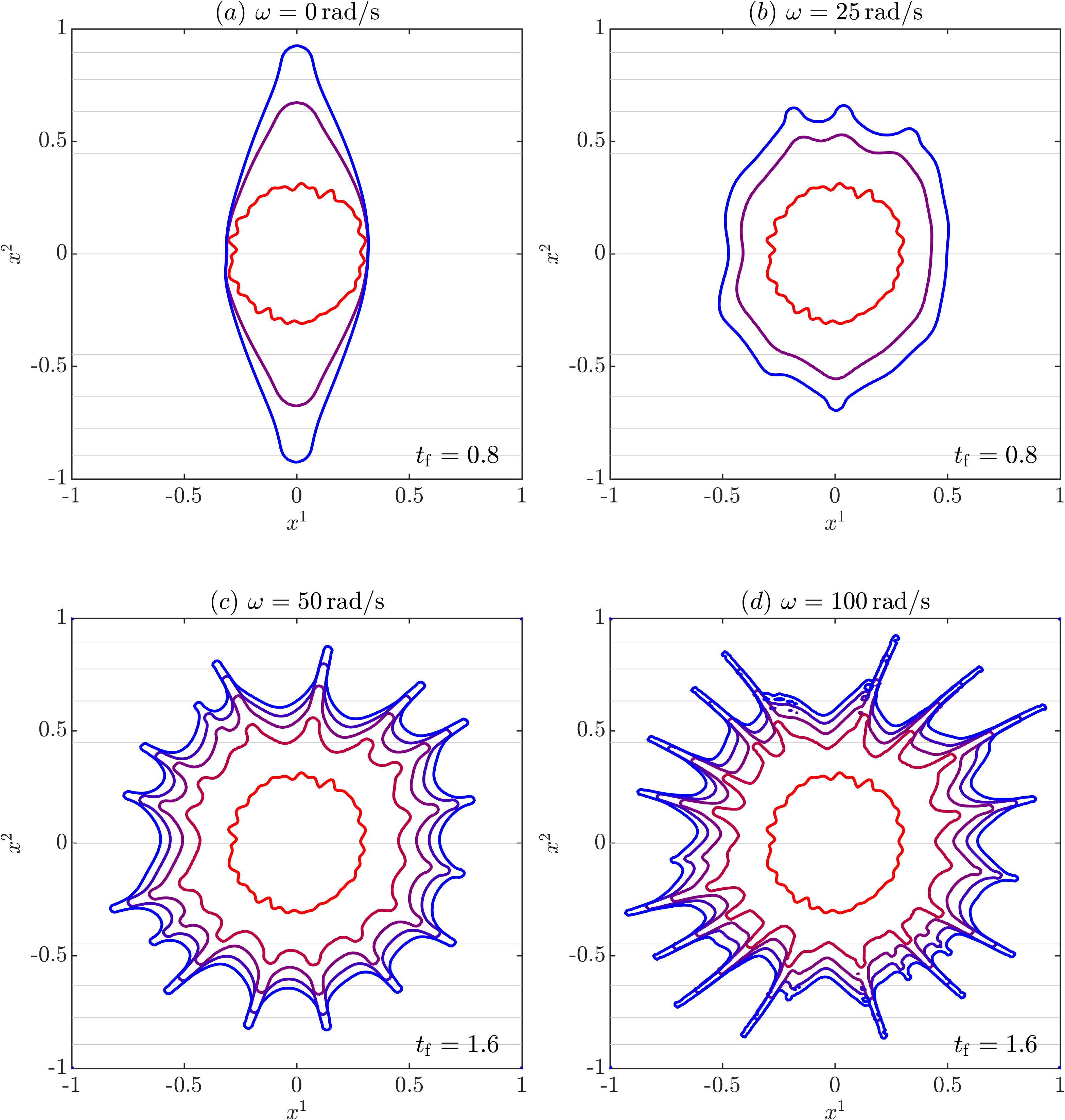}
    \caption{Evolution of the contact line on a parabolic cylinder substrate (equation (\ref{eq:substrate geometry cylinder}) and figure \ref{fig:Substrate geometry}a) from a randomly perturbed initial condition in intervals of $\Delta t = 0.4$, coloured from red to blue with increasing $t$ up to $t_\text{f}$ on an anticlockwise-rotating substrate.
    Substrate contours with vertical spacing $\Delta z = 0.1$ are shown in grey.
    (\textit{a}) Contact line up to $t_\text{f}=0.8$ on a stationary substrate.
    (\textit{b}) Contact line up to $t_\text{f}=1$ with $\omega = 25\,\text{rad/s}$.
    (\textit{c}) Contact line up to $t_\text{f}=1.6$ with $\omega = 50\,\text{rad/s}$. 
    (\textit{d}) Contact line up to $t_\text{f}=1.6$ with $\omega = 100\,\text{rad/s}$.}
    \label{fig:Final contact line cylinder perturbation}
\end{figure}

Figure \ref{fig:Final contact line cylinder perturbation} shows the evolution of the contact line of a perturbed spherical droplet (\ref{eq:perturbation initial condition}) on a rotating parabolic cylinder substrate for angular velocities from $\omega = 0\,\text{rad/s}$ to $100\,\text{rad/s}$.
The substrate is described in dimensionless Cartesian coordinates by
\begin{equation}\label{eq:substrate geometry cylinder}
    \vect{s}(x^1,x^2) = \bigg[ x^1, x^2, - \frac{(x^2)^2}{2} \bigg],
\end{equation}
and shown in figure \ref{fig:Substrate geometry}\textit{a}.
In the gravity-driven regime on a stationary substrate ($\omega = 0\,\text{rad/s}$, figure \ref{fig:Final contact line cylinder perturbation}\textit{a}), the initial droplet forms rivulets flowing down either side of the substrate in the direction of $\pm x^2$.
As the angular velocity of the substrate is increased, the film dynamics enter a transitional regime, where there are competing effects from gravitational and centrifugal forces.
At $\omega = 25\,\text{rad/s}$ (figure \ref{fig:Final contact line cylinder perturbation}\textit{b}), the droplet begins to spread radially in all directions due to centrifugal force, but still with a preference toward the $\pm x^2$ directions and forming rivulets down either side of the substrate due to gravity.
As the angular velocity is increased further, the dynamics become driven primarily by centrifugal force.
At $\omega = 50\,\text{rad/s}$ and $100\,\text{rad/s}$ (figures \ref{fig:Final contact line cylinder perturbation}\textit{c} and \ref{fig:Final contact line cylinder perturbation}\textit{d}), there is weak gravitational influence and the droplet spreads evenly in all directions.
In this regime, the spreading droplet develops a fingering instability at the contact line similarly to on a flat substrate, with the growth rate of the fingers increasing with increasing angular velocity.
At the leading order, the volume flux over the substrate surface (\ref{eq:general volume flux components}) is driven by the component of the total body force tangent to the substrate surface.
In the centrifugal force-driven regime, the tangential component of the total body force is greatest along the ridge of the substrate.
At high angular velocity ($\omega = 100\,\text{rad/s}$, figure \ref{fig:Final contact line cylinder perturbation}\textit{d}), this leads to a deflection of fingers in the $\pm x^1$ directions.

\subsubsection{Saddle substrate}\label{sec:saddle results}
\begin{figure}
    \centering
    \includegraphics{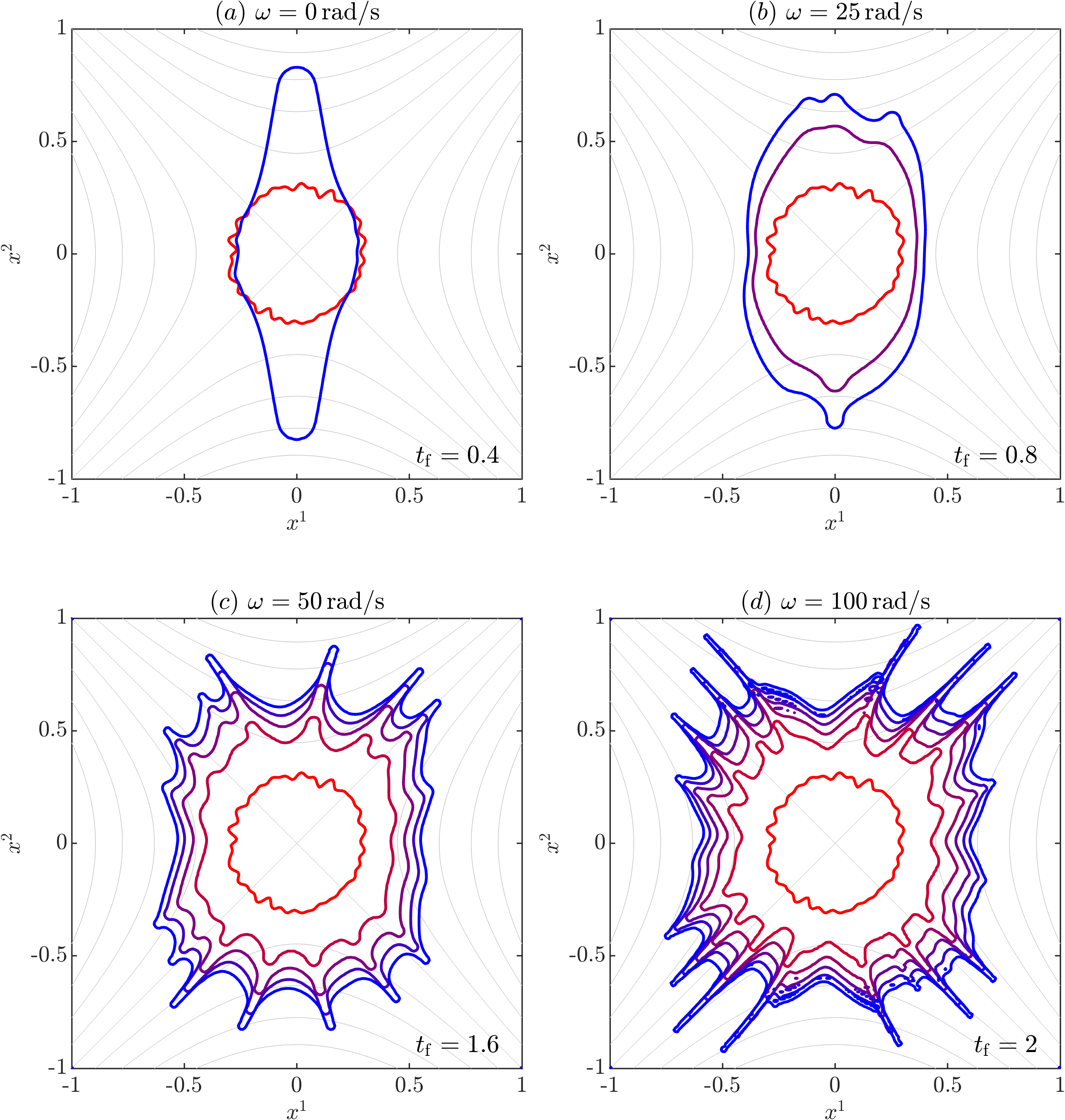}
    \caption{Evolution of the contact line on a saddle substrate (equation (\ref{eq:substrate geometry saddle}) and figure \ref{fig:Substrate geometry}b) from a randomly perturbed initial condition in intervals of $\Delta t = 0.4$, coloured from red to blue with increasing $t$ up to $t_\text{f}$ on an anticlockwise-rotating substrate.
    Substrate contours with vertical spacing $\Delta z = 0.1$ are shown in grey.
    (\textit{a}) Contact line up to $t_\text{f}=0.4$ on a stationary substrate.
    (\textit{b}) Contact line up to $t_\text{f}=1$ with $\omega = 25\,\text{rad/s}$.
    (\textit{c}) Contact line up to $t_\text{f}=1.6$ with $\omega = 50\,\text{rad/s}$. 
    (\textit{d}) Contact line up to $t_\text{f}=2$ with $\omega = 100\,\text{rad/s}$.}
    \label{fig:Final contact line saddle perturbation}
\end{figure}

Figure \ref{fig:Final contact line saddle perturbation} shows the evolution of the contact line of a perturbed spherical droplet (\ref{eq:perturbation initial condition}) on a rotating saddle substrate for angular velocities from $\omega = 0\,\text{rad/s}$ to $100\,\text{rad/s}$.
The substrate is described in dimensionless Cartesian coordinates by
\begin{equation}\label{eq:substrate geometry saddle}
    \vect{s}(x^1,x^2) = \bigg[ x^1, x^2, \frac{(x^1)^2}{2} - \frac{(x^2)^2}{2} \bigg],
\end{equation}
and shown in figure \ref{fig:Substrate geometry}\textit{b}.
Similarly to the parabolic cylinder, in the gravity-driven regime ($\omega = 0\,\text{rad/s}$, figure \ref{fig:Final contact line saddle perturbation}\textit{a}), the initial droplet forms rivulets  along the downward-sloping $\pm x^2$ directions.
At $\omega = 25\,\text{rad/s}$ (figure \ref{fig:Final contact line saddle perturbation}\textit{b}), the droplet again begins to spread radially in all directions, while still forming rivulets in the downward-sloping directions.
On a saddle substrate, however, the centrifugal force is not strong enough to overcome the upward slope of the substrate in the $\pm x^1$ directions, leading to a more pronounced elliptically-shaped contact line.
At high angular velocities, the film dynamics on a saddle substrate begin to differ significantly from those on the parabolic cylinder.
In particular, for centrifugal force-driven flow on a saddle substrate, the tangential component of the total body force is greatest on the diagonals $x^1 = \pm x^2$, along which the substrate is horizontal.
As with the parabolic cylinder substrate, we continue to observe that the thickness of the film behind the moving front is the same as on a flat substrate at high angular velocity ($\omega = 100\,\text{rad/s}$).
At $\omega = 50\,\text{rad/s}$ and $100\,\text{rad/s}$ (figures \ref{fig:Final contact line cylinder perturbation}\textit{c} and \ref{fig:Final contact line cylinder perturbation}\textit{d}), the spreading droplet again develops a fingering instability at the contact line.
In this case, however, the fingering does not develop in all directions, and grows in an `X'-shape, primarily along the diagonal $z=0$ contours ($x^1 = \pm x^2$) in the direction of the greatest tangential body force.

\subsection{Film thickness and surface coverage}\label{sec:film thickness and coverage}
\begin{figure}
\centering
\includegraphics{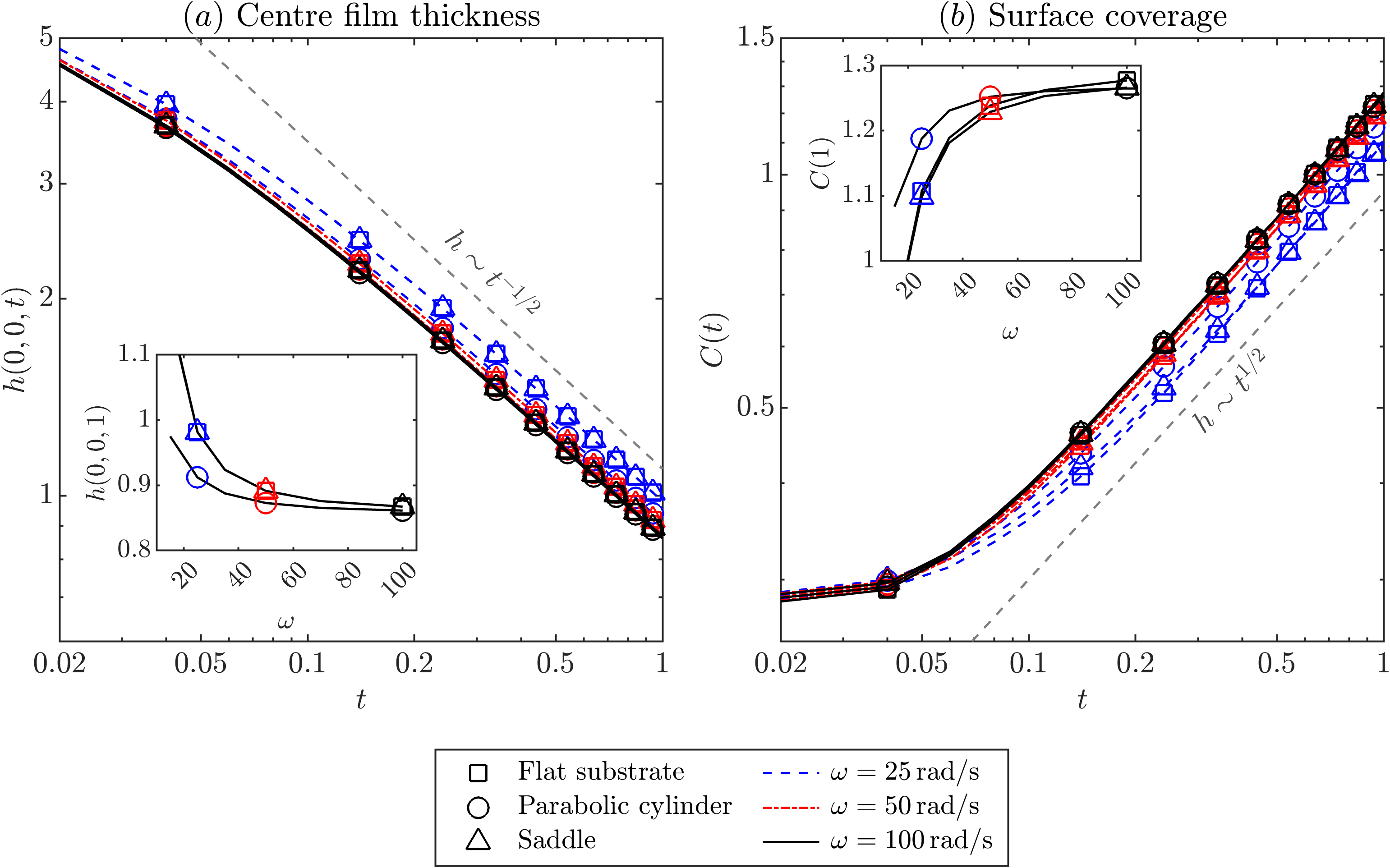}
\caption{
Temporal evolution of $(a)$~the film thickness at the substrate centre, $h(0,0,t)$, and $(b)$~the surface area of the fluid-covered region, $C(t)$, on flat, parabolic cylinder and saddle substrates for  angular velocities $\omega=25\,\text{rad/s}$, $50\,\text{rad/s}$ and $100\,\text{rad/s}$.
In each case, the inset shows the film thickness or surface coverage at $t=1$ as a function of the substrate angular velocity, $\omega$.
The lines between markers in the inset plots include results for intermediate angular velocities not shown in the main plot.
}
\label{fig:Film_Thickness_Surface_Coverage}
\end{figure}

To quantify the film evolution, we report in figure~\ref{fig:Film_Thickness_Surface_Coverage}a the film thickness at the substrate centre, $h(0,0,t)$, on the three considered substrates (flat, parabolic cylinder and saddle), for several values of the angular velocity.
On all substrates and for all angular velocities, the centre film thickness decreases as $h \sim t^{-1/2}$ from $t \approx 0.1$ onwards, and the variations due to the substrate and angular velocity are small compared to the overall variation over time.
Furthermore, at a high angular velocity ($\omega = 100\,\text{rad/s}$), the centre film thickness remains the same as on a flat substrate while the droplet spreads.
The same power law for film thickness evolution on cylindrical and spherical substrates without rotation \citep{Takagi2010} and flat substrates with rotation \citep{Melo1989} was measured experimentally and predicted analytically by self-similar solutions derived from the lubrication equation with surface tension and hydrostatic pressure neglected away from the contact line.

We also define, and report in figure~\ref{fig:Film_Thickness_Surface_Coverage}b, the surface coverage at time $t$ as the substrate surface area within the wetted region,
\begin{equation}
    C(t) = \iint_\Omega \phi(x,^1,x^2,t) \sqrt{ G(x^1,x^2) } \,\mathrm{d}x^1\mathrm{d}x^2,
\end{equation}
where $\Omega = [-1,1]\times[-1,1]$ is the domain and
\begin{equation}
    \phi(x,^1,x^2,t) = \begin{cases}
        1 & \text{if }h(x^1,x^2,t)\geq 5 h_\text{p}, \\
        0 & \text{otherwise.}
    \end{cases}
\end{equation}
Here, we again choose $h \geq 5 h_\text{p}$ as the threshold for the wetted area to be consistent with the contact lines plotted in sections \ref{sec:flat substrate} and \ref{sec:non-axisymmetric substrate}.

As expected, larger angular velocities lead to faster initial film thinning and, therefore, to thinner films and larger surface coverage at later times, scaling as $C\sim t^{1/2}$.
At smaller angular velocities (gravity-driven flows), thinner films are obtained on the parabolic cylinder substrate than on the flat and saddle substrates, which may be ascribed to the larger substrate curvature at the centre.
At larger angular velocities (centrifugal force-driven flows), however, the substrate geometry does not significantly affect the overall rate of dispersal over the surface.
The key difference in the results between different geometries is the shape of the droplet: although the film spreads a similar amount, its distribution is no longer axisymmetric on the parabolic cylinder and saddle substrates.

\section{Conclusions}
In this work, we have derived and implemented a model for the evolution of a thin fluid film on a rotating curved substrate surface, allowing for an arbitrary substrate parameterisation, and including the effects of surface tension, disjoining pressure, gravity, centrifugal, and Coriolis forces.
We can consider this either as an extension of existing models of spin coating (such as \citet{Schwartz2004} and \citet{Weidner2018}) to allow for more complex and non-axisymmetric substrate geometry, or as an extension of models for thin film flow over stationary curved substrates (such as \citet{Roy2002}, \citet{Thiffeault2006}, and \citet{Mayo2015}) to include centrifugal and Coriolis forces due to substrate rotation.
We implemented our generalised model using COMSOL Multiphysics and MATLAB to simulate the evolution of spin-coated films on flat and curved substrates at a range of angular velocities.

On a flat substrate, we demonstrated that the Coriolis force has a negligible effect at low substrate angular velocity, and at high angular velocity leads to a slight deflection of the flow against the direction of substrate rotation.
This deflection was observed in both the instantaneous initial volume flux at the start of spin coating, and in the fingering instability at the contact line as the spin-coated droplet spreads.
Our analysis agrees with the conclusions from \citet{Myers2001}, showing that while the Coriolis force has only a small effect, there is a range of parameters (when $\mathit{Ta} \sim 1$) where the Coriolis force cannot be neglected, but where inertial effects can still be ignored.

On parabolic cylinder and saddle substrate geometries, we showed example simulations of spin coating at a range of angular velocities to demonstrate the transition from gravity-driven to centrifugal force-driven flow.
In both cases, as the angular velocity is increased, a droplet initially at the centre of the substrate shifted from draining in the downward-sloping directions to spreading in the directions where the tangential component of the centrifugal force is greatest.
On a parabolic cylinder substrate at $\omega = 100\,\text{rad/s}$, this leads to a similar pattern of fingering at the contact line to a flat substrate, however with the fingers deflected along the direction of the ridge (the $\pm x^1$ directions).
On a saddle substrate at $\omega = 100\,\text{rad/s}$, the effect of substrate geometry was clearer, with the fingers at the contact line growing in an `X'-shape (along the $x^1 = \pm x^2$ diagonals).
Furthermore, we showed that for both parabolic cylinder and saddle substrates, the film thickness at the substrate centre and the wetted area remain similar over time to those for a droplet spreading on a flat substrate, demonstrating that the key difference between substrate geometries is the shape of the spreading droplet and the direction of fingering.

There are several natural ways in which the present work could be extended.
In particular, in order to investigate higher angular velocities where the Coriolis force would have a stronger effect (where $\mathit{Ta} > 1$, and hence $\mathit{Re} > \varepsilon$), however, a model incorporating inertial effects would be required.
Additionally, a stability analysis around the contact line on a curved substrate with both gravitational and centrifugal forces may provide a more detailed explanation for the complex fingering patterns observed on a saddle substrate.

\backsection[Funding]{This work is part of the project "Development of a multi-axis spin-coating system to coat curved surfaces" funded by the New Zealand Ministry of Business, Innovation and Employment Endeavour fund (grant number UOCX1904). This funding is gratefully acknowledged.}

\backsection[Declaration of interests]{The authors report no conflict of interest.}

\backsection[Author ORCIDs]{Ross G.~Shepherd, https://orcid.org/0000-0003-3489-018X; Edouard Boujo, https://orcid.org/0000-0002-4448-6140; Mathieu Sellier, https://orcid.org/0000-0002-5060-1707}

\appendix
\section{Integration of the Navier--Stokes equations normal to the substrate}\label{app:integration of NS equations}
% Define perturbation expansion
In order to integrate the simplified NS equations (\ref{eq:NS lubrication}) in the direction normal to the substrate, we introduce an perturbation expansion in $\varepsilon$ for the velocity, pressure, and body force as follows, and omit the tildes over dimensionless variables for brevity:
\begin{equation}
\arraycolsep=1.5pt
\left. \begin{array}{ccccccl}
u^\alpha &=& u^\alpha_{(0)} &+& \varepsilon u^\alpha_{(1)} &+& O(\varepsilon^2), \\[4pt]
p &=& p_{(0)} &+& \varepsilon p_{(1)} &+& O(\varepsilon^2), \\[4pt]
f^\alpha &=&  f^\alpha_{(0)} &+& \varepsilon f^\alpha_{(1)} &+& O(\varepsilon^2), \\[4pt]
f^n &=& f^n_{(0)} &+& O(\varepsilon).
\end{array}\quad\right\} \label{eq:perturbation expansion u p and f}
\end{equation}
Substituting (\ref{eq:basis vectors dimensionless}) for $\vect{e}^{+\alpha}$, the $O(1)$ and $O(\varepsilon)$ parts of the tangential body force can be written as
\begin{equation}
f^\alpha_{(0)} = \big[
N_\text{grav} \hat{\vect{g}}
- N_\text{cent} \hat{\vect{\omega}} \times \big( \hat{\vect{\omega}} \times \vect{s} )
\big] \vect{\cdot} \vect{e}^{\alpha},
\end{equation}
\begin{equation}\label{eq:f alpha 1}
f^\alpha_{(1)} = - \big[ n N_\text{cent} \hat{\vect{\omega}} \times ( \hat{\vect{\omega}} \times \hat{\vect{n}} )
+ 2 \mathit{Ta} \hat{\vect{\omega}} \times \vect{u}_{(0)}
\big] \vect{\cdot} \vect{e}^\alpha
+ n K_\beta{}^\alpha f^\beta_{(0)},
\end{equation}
where $\vect{u}_{(0)} = u^\alpha_{(0)} \vect{e}_\alpha$ is the leading-order velocity vector.
Similarly, the leading-order normal body force is
\begin{equation}
f^n_{(0)} = \big[
N_\text{grav} \hat{\vect{g}}
- N_\text{cent} \hat{\vect{\omega}} \times \big( \hat{\vect{\omega}} \times \vect{s} )
\big] \vect{\cdot} \hat{\vect{n}}.
\end{equation}
Here, we see that the $O(1)$ body force terms are constant in $n$, but the $O(\varepsilon)$ terms depend on $n$ both explicitly and implicitly due to variation in $\vect{u}_{(0)}$ in the normal direction.

% O(1) and O(epsilon) lubrication equations
Substituting the perturbation expansion (\ref{eq:perturbation expansion u p and f}), together with (\ref{eq:basis vectors dimensionless}) for $G^{+\alpha\beta}$, the lubrication equations (\ref{eq:NS lubrication}) at $O(1)$ and $O(\varepsilon)$ are
\begin{equation}\label{eq:NS order 1}
\left. \begin{array}{c}
\displaystyle{
-\frac{\partial p_{(0)}}{\partial n} = 0, }\\[8pt]
\displaystyle{
-\nabla^\alpha p_{(0)}
+ \frac{\partial^2 u^\alpha_{(0)}}{\partial n^2}
+ f^\alpha_{(0)} = 0, }
\end{array} \quad\right\}
\end{equation}
\begin{equation}\label{eq:NS order epsilon}
\left. \begin{array}{c}
\displaystyle{
-\frac{\partial p_{(1)}}{\partial n} + f^n_{(0)} = 0, }\\[8pt]
\displaystyle{
-\nabla^\alpha p_{(1)}
-2 n K_\beta{}^\alpha \nabla^\beta p_{(0)}
+ \frac{\partial^2 u^\alpha_{(1)}}{\partial n^2}
- \big( \kappa \delta_\beta{}^\alpha + 2 K_\beta{}^\alpha \big) \frac{\partial u^\beta_{(0)}}{\partial n}
+ f^\alpha_{(1)} = 0, }
\end{array} \quad\right\}
\end{equation}
where $\nabla^\alpha = G^{\alpha\beta} \partial/\partial x^\beta$ is a contravariant component of the substrate gradient.
The corresponding perturbation expansions of the dimensionless boundary conditions are ${u^\alpha_{(0)} = u^\alpha_{(1)} = 0}$ on $n=0$, and
\begin{equation}\label{eq:perturbation boundary conditions}
\left. \begin{array}{cc}
p_{(0)} = -N_\text{surf}\kappa - \varPi, &
p_{(1)} = -N_\text{surf}(\kappa_2 h + \Delta_S h), \\[4pt]
\displaystyle\frac{\partial u^\alpha_{(0)}}{\partial n} = 0, &
\displaystyle\frac{\partial u^\alpha_{(1)}}{\partial n} = 0,
\end{array} \quad\right\} \quad \text{on }n=h.
\end{equation}

% Integrate O(1) terms
Integrating (\ref{eq:NS order 1}) with respect to $n$ with boundary conditions (\ref{eq:perturbation boundary conditions}) gives the standard leading-order lubrication model:
\begin{equation}\label{eq:leading order velocity}
p_{(0)} = -N_\text{surf}\kappa - \varPi, \qquad
u^\alpha_{(0)} =
\bigg( hn - \frac{n^2}{2} \bigg)
\bigg( \nabla^\alpha (N_\text{surf}\kappa + \varPi) + f^\alpha_{(0)} \bigg),
\end{equation}
which is equivalent to equation (34) from \citet{Roy2002} and equation (IV.7) from \citet{Thiffeault2006}.

Let $\omega^\alpha = \hat{\vect{\omega}} \vect{\cdot} \vect{e}^\alpha$ and $\omega^n = \hat{\vect{\omega}} \vect{\cdot} \hat{\vect{n}}$ be the tangential and normal components of $\hat{\vect{\omega}}$ in the substrate coordinate system, and let $\epsilon_\beta{}^\alpha$ be a mixed component of the modified Levi--Civita tensor, $\tens{E} = \epsilon_{\alpha\beta} \vect{e}^\alpha \vect{e}^\beta$, defined by
\begin{equation}\label{eq:levi-civita symbol appendix}
\epsilon_{\alpha\beta}
= ( \vect{e}_\alpha \times \vect{e}_\beta ) \vect{\cdot} \hat{\vect{n}}
= \big\lVert \vect{e}_1 \times \vect{e}_2 \big\rVert
\begin{cases}
1 & \text{if }\alpha=1,\,\beta=2, \\
-1 & \text{if }\alpha = 2,\,\beta=1, \\
0 & \text{otherwise.}
\end{cases}
\end{equation}
This allows the Coriolis force term in (\ref{eq:f alpha 1}) to be written as
\begin{equation}
    ( \hat{\vect{\omega}} \times \vect{u}_{(0)} ) \vect{\cdot} \vect{e}^\alpha
    = \omega^n \epsilon_\beta{}^\alpha \bigg( hn - \frac{n^2}{2} \bigg)
    \bigg( \nabla^\beta (N_\text{surf}\kappa + \varPi) + f^\beta_{(0)} \bigg).
\end{equation}
The $O(\varepsilon)$ tangential body force can then be expressed explicitly in terms of $n$:
\begin{equation}\label{eq:f alpha 1 explicit}
\begin{split}
f^\alpha_{(1)} = &\; n N_\text{cent} ( \hat{\vect{\omega}} \times ( \hat{\vect{\omega}} \times \hat{\vect{n}} )) \vect{\cdot} \vect{e}^\alpha
+ n K_\beta{}^\alpha f^\beta_{(0)} \\
& - 2 \mathit{Ta} \omega^n \epsilon_\beta{}^\alpha \bigg( hn - \frac{n^2}{2} \bigg)
\bigg( \nabla^\beta (N_\text{surf}\kappa + \varPi) + f^\beta_{(0)} \bigg).
\end{split}
\end{equation}
Integrating the pressure equation from (\ref{eq:NS order epsilon}) in $n$ with the boundary condition (\ref{eq:perturbation boundary conditions}) gives
\begin{equation}\label{eq:order epsilon pressure}
p_{(1)} = -N_\text{surf}(\kappa_2 h + \Delta_S h) - f^n_{(0)}(h-n).
\end{equation}
Now substituting (\ref{eq:f alpha 1 explicit}) and (\ref{eq:order epsilon pressure}) and collecting like terms, we can re-write (\ref{eq:NS order epsilon}) as
\begin{equation}\label{eq:NS order epsilon ABC}
    A^\alpha + n B^\alpha + n^2 C^\alpha + \frac{\partial^2 u^\alpha_{(1)}}{\partial n^2} = 0,
\end{equation}
where
\begin{equation}\label{eq:ABC}
\arraycolsep=1.5pt
\left. \begin{array}{rl}
    A^\alpha = &\; N_\text{surf} \nabla^\alpha (k_2 h + \Delta_S h) + \nabla^\alpha \big(h f^n_{(0)} \big) \\[4pt]
    & -\; h \big( \kappa \delta_\beta{}^\alpha + 2 K_\beta{}^\alpha \big) \big( \nabla^\beta (N_\text{surf}\kappa + \varPi) + f^\beta_{(0)} \big),\\[8pt]
    B^\alpha = &-\; \nabla^\alpha f^n_{(0)}
    +\; N_\text{cent} ( \hat{\vect{\omega}} \times ( \hat{\vect{\omega}} \times \hat{\vect{n}} )) \vect{\cdot} \vect{e}^\alpha \\[4pt]
    &+\; \big( \kappa \delta_\beta{}^\alpha + 4 K_\beta{}^\alpha - 2\mathit{Ta} h \omega^n \epsilon_\beta{}^\alpha \big) \nabla^\beta (N_\text{surf}\kappa + \varPi) \\[4pt]
    &+\; \big( \kappa \delta_\beta{}^\alpha + 3 K_\beta{}^\alpha - 2\mathit{Ta} h \omega^n \epsilon_\beta{}^\alpha \big) f^\beta_{(0)},\\[8pt]
    C^\alpha = & 2\mathit{Ta} \omega^n \epsilon_\beta{}^\alpha \big( \nabla^\beta (N_\text{surf}\kappa + \varPi) + f^\beta_{(0)} \big).
\end{array}\quad\right\}
\end{equation}
Integrating (\ref{eq:NS order epsilon ABC}) with boundary conditions (\ref{eq:perturbation boundary conditions}), the $O(\varepsilon)$ velocity is
\begin{equation}
u^\alpha_{(1)} = A^\alpha \bigg(hn-\frac{n^2}{2}\bigg) + B^\alpha \bigg( \frac{h^2 n}{2} - \frac{n^3}{6} \bigg) + C^\alpha \bigg( \frac{h^3 n}{3} - \frac{n^4}{12} \bigg).
\end{equation}

% Perturbation expansion of q
Similar to (\ref{eq:perturbation expansion u p and f}), we introduce a perturbation expansion for the dimensionless volume flux over the substrate surface:
\begin{equation}\label{eq:perturbation expansion q}
q^\alpha = q^\alpha_{(0)} + \varepsilon q^\alpha_{(1)} + O(\varepsilon^2).
\end{equation}
Substituting (\ref{eq:perturbation expansion u p and f}) into (\ref{eq:dimensionless volume flux}), the $O(1)$ and $O(\varepsilon)$ terms of the flux are
\begin{equation}
q^\alpha_{(0)} = \int_0^h u^\alpha_{(0)} \, \mathrm{d} n,
\end{equation}
\begin{equation}
q^\alpha_{(1)} = \int_0^h \big( u^\alpha_{(1)} - \kappa n u^\alpha_{(0)} \big) \mathrm{d} n.
\end{equation}
Integrating, the leading-order flux can be expressed as
\begin{equation}\label{eq:order 1 volume flux}
q^\alpha_{(0)} = \frac{h^3}{3} \bigg( \nabla^\alpha (N_\text{surf}\kappa + \varPi) + f^\alpha_{(0)} \bigg),
\end{equation}
and the $O(\varepsilon)$ flux as
\begin{equation}\label{eq:order epsilon volume flux}
q^\alpha_{(1)} = \frac{h^3}{3} A^\alpha + \frac{5 h^4}{24} \bigg[ B^\alpha - \kappa \bigg( \nabla^\alpha (N_\text{surf}\kappa + \varPi) + f^\alpha_{(0)} \bigg) \bigg] + \frac{3 h^5}{20} C^\alpha.
\end{equation}
Finally, substituting (\ref{eq:ABC}) and combining the $O(1)$ and $O(\varepsilon)$ terms, the components of the dimensionless volume flux over the substrate are
\begin{equation}\label{eq:general volume flux components appendix}
\begin{split}
q^\alpha
= \frac{h^3}{3} \bigg[ & \bigg( \delta_\beta{}^\alpha - \varepsilon h \bigg( \kappa \delta_\beta{}^\alpha - \frac{1}{2} K_\beta{}^\alpha \bigg) - \varepsilon h^2 \omega^n \frac{4 \mathit{Ta}}{5} \epsilon_\beta{}^\alpha \bigg) \nabla^\beta (N_\text{surf}\kappa^* + \varPi) \\
& + \bigg( \delta_\beta{}^\alpha - \varepsilon h \bigg( \kappa \delta_\beta{}^\alpha + \frac{1}{2} K_\beta{}^\alpha \bigg) - \varepsilon h^2 \omega^n \frac{4 \mathit{Ta}}{5} \epsilon_\beta{}^\alpha \bigg) f^\beta_{(0)} \\
& + \varepsilon f^n_{(0)} \nabla^\alpha h 
+ \varepsilon h N_\text{cent} ( \hat{\vect{\omega}} \times ( \hat{\vect{\omega}} \times \hat{\vect{n}} )) \vect{\cdot} \vect{e}^\alpha
\bigg]
+ O(\varepsilon\mathit{Re}, \varepsilon^2).
\end{split}
\end{equation}

\section{Effect of precursor film thickness}\label{app:precursor}
\begin{figure}
\centerline{ 
    \begin{overpic}[width=0.49\textwidth, trim=0mm 0mm 0mm 0mm, clip=true]{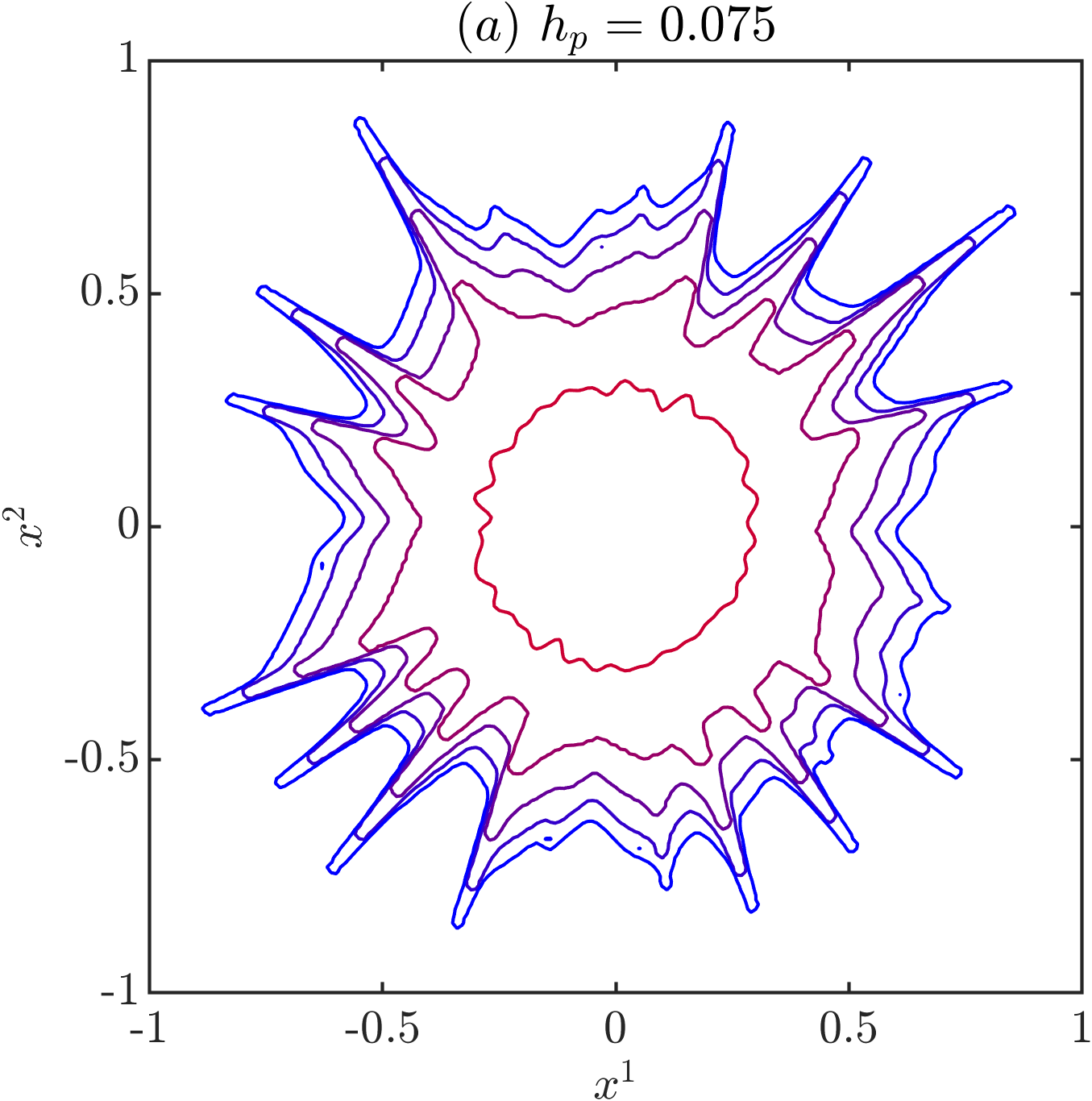}  
 	\end{overpic}
    \begin{overpic}[width=0.49\textwidth, trim=0mm 0mm 0mm 0mm, clip=true]{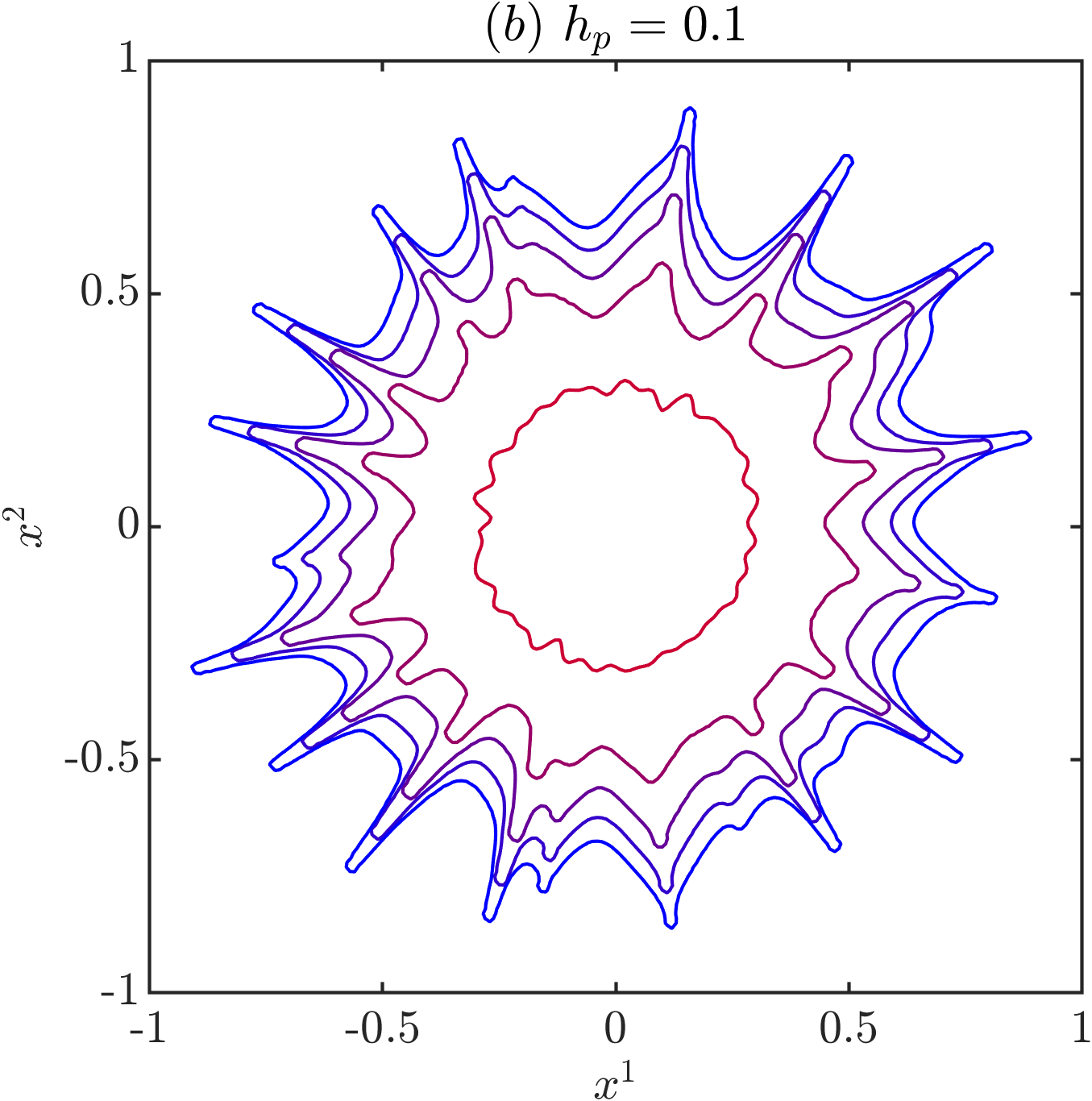}  
 	\end{overpic}
}
\vspace{0.7cm}
\centerline{ 
    \begin{overpic}[width=0.49\textwidth, trim=0mm 0mm 0mm 0mm, clip=true]{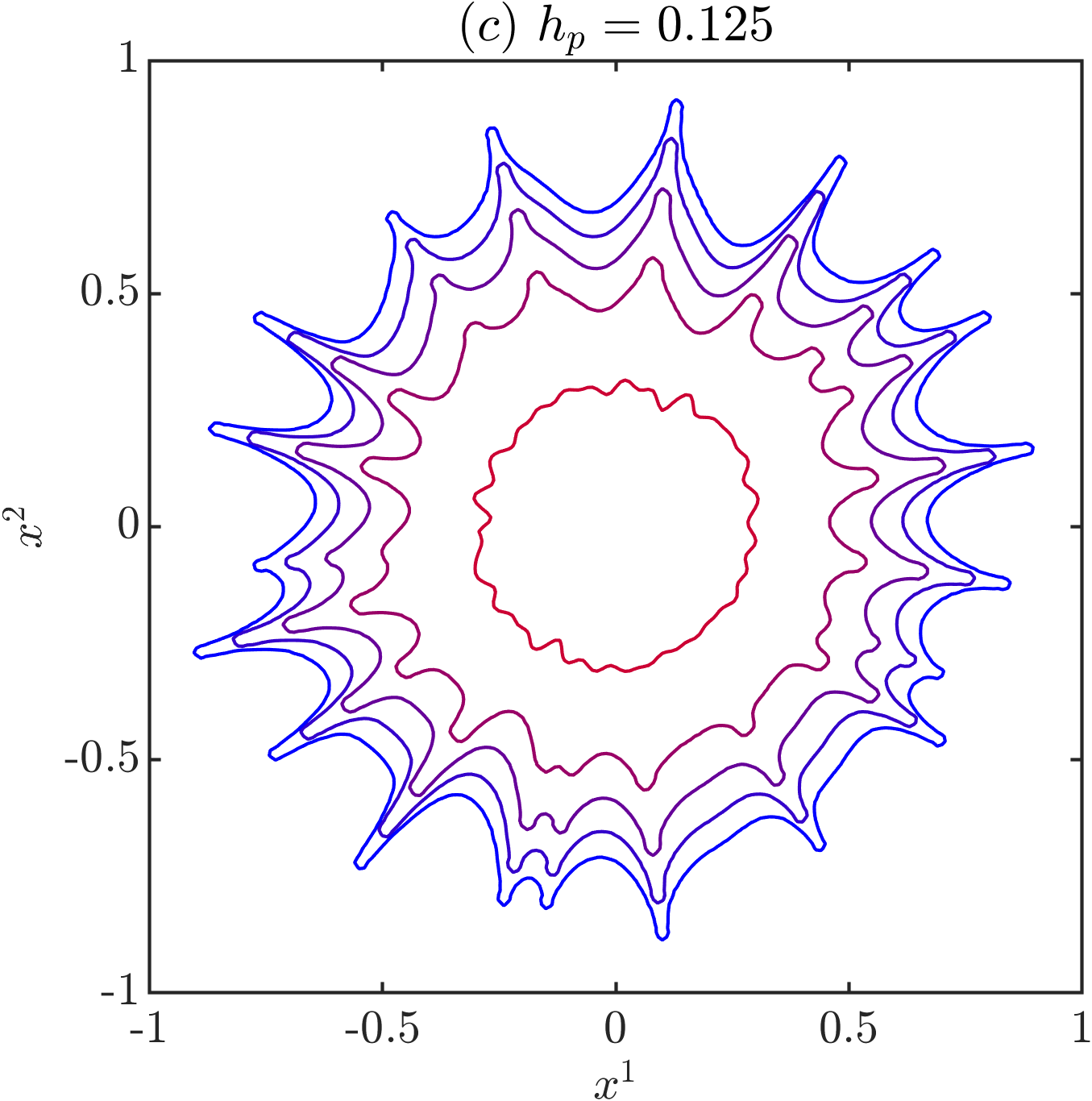}  
 	\end{overpic}
    \begin{overpic}[width=0.49\textwidth, trim=0mm 0mm 0mm 0mm, clip=true]{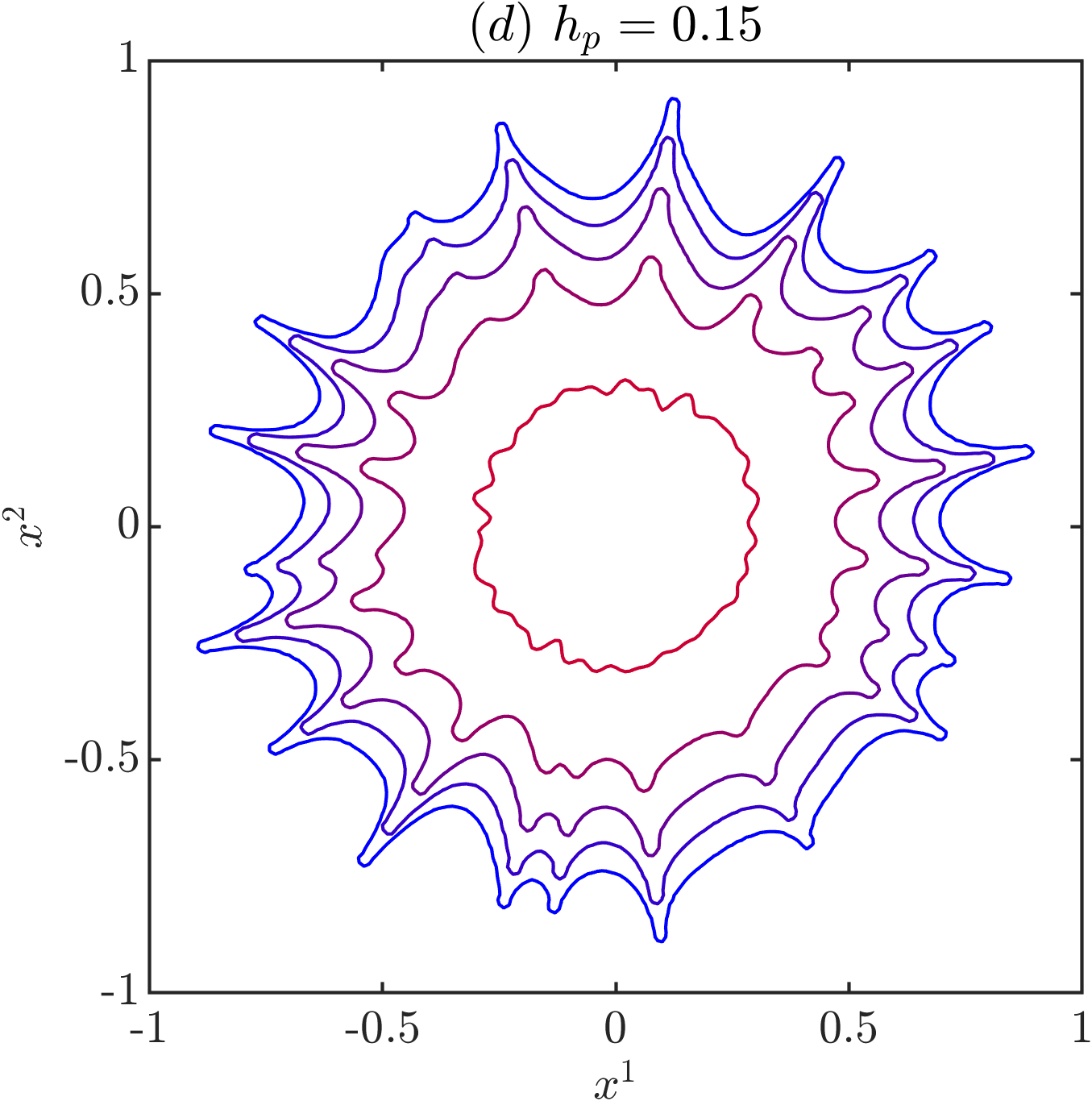}  
 	\end{overpic}
}
\caption{
Evolution of the contact line on a flat substrate from a randomly perturbed initial condition in intervals of $\Delta t = 0.4$, coloured from red to blue with increasing $t$ up to $t_\text{f}=1.6$, on an anticlockwise-rotating substrate with $\omega = 100\,\text{rad/s}$, for different values of the precursor film thickness: 
$(a)$~$h_p = 0.075$,
$(b)$~$h_p = 0.1$,
$(c)$~$h_p = 0.125$,
$(d)$~$h_p = 0.15$.   
}
\label{fig:hp}
\end{figure}

Figure~\ref{fig:hp} shows the effect of the precursor film thickness on the evolution of the contact line on the flat substrate rotating at $\omega = 100\,\text{rad/s}$. 
Differences are observed when $h_p \leq 0.075$, which are due to a numerical instability developing when the mesh is too coarse compared to the precursor film thickness.
Conversely, the results are converged and robust for $h_p \geq 0.1$: while small localised differences can be observed, the overall spreading rate and shape are unaffected. In any case, the conclusions of the present study obtained with $h_p=0.1$ remain unchanged.

\section{Effect of the initial condition}\label{app:random}
\begin{figure}
\centerline{ 
    \begin{overpic}[width=0.49\textwidth, trim=0mm 0mm 0mm 0mm, clip=true]{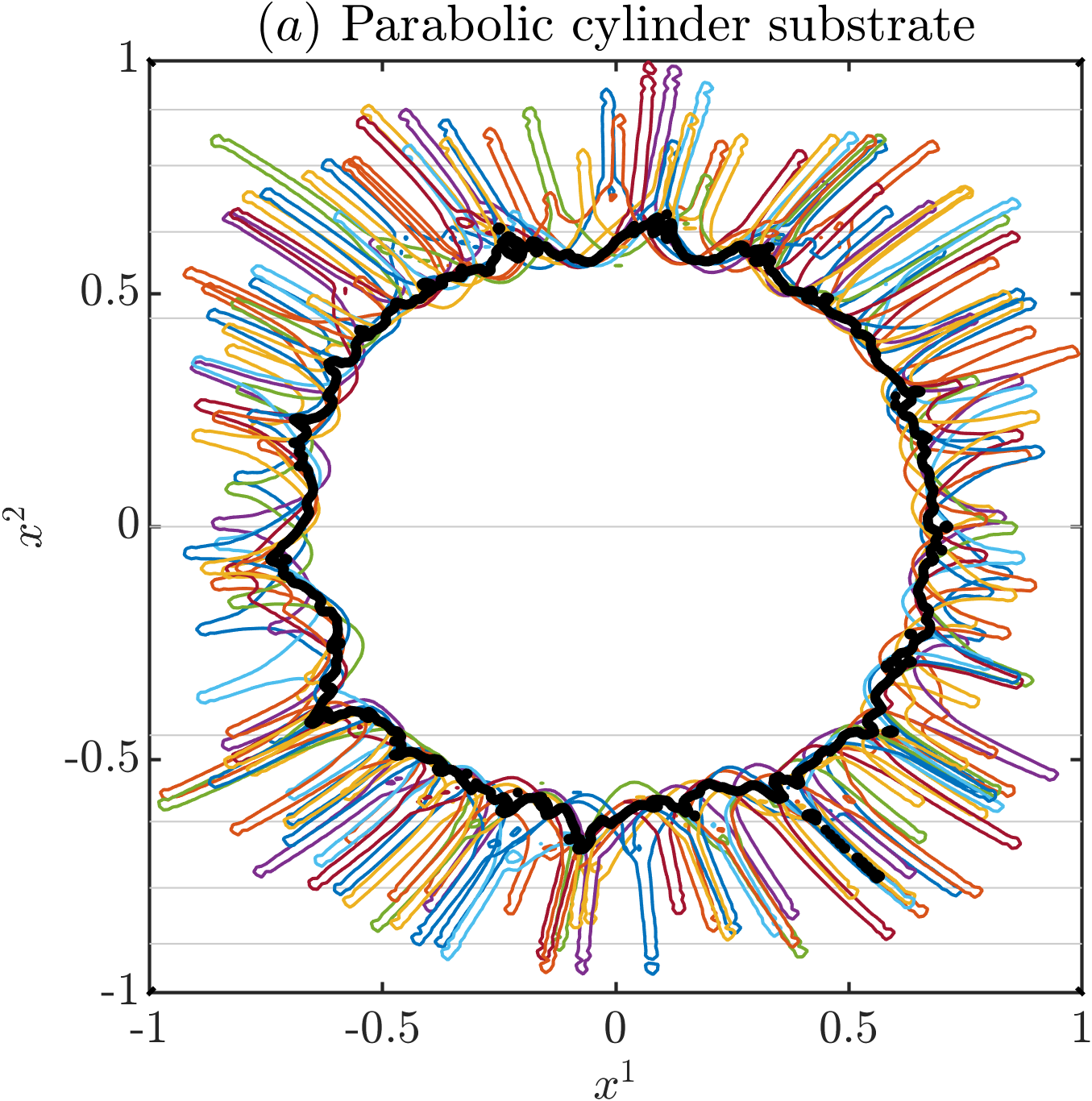}  
 	\end{overpic}
    \begin{overpic}[width=0.49\textwidth, trim=0mm 0mm 0mm 0mm, clip=true]{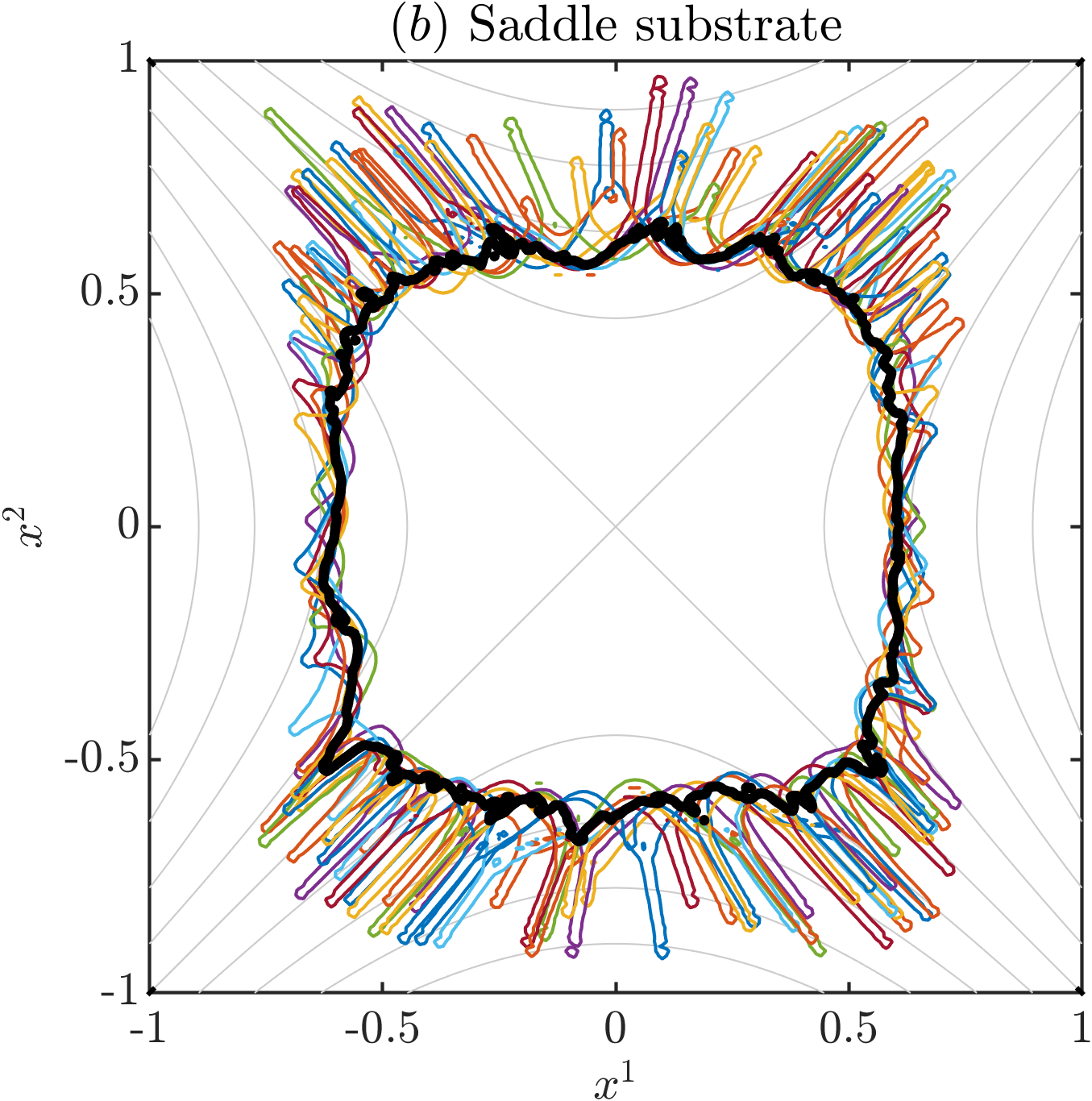}  
 	\end{overpic}
}
\caption{
Contact line at $t=1.6$ from randomly perturbed initial conditions on an anticlockwise-rotating $(a)$ parabolic cylinder substrate and $(b)$ saddle substrate with $\omega = 100\,\text{rad/s}$.
Thin coloured lines: 10 independent realisations.
Thick black line: ensemble-averaged film.
Substrate contours with vertical spacing $\Delta z = 0.1$ are shown in grey.
}
\label{fig:random}
\end{figure}

Figure~\ref{fig:random} shows the contact lines obtained at $t=1.6$ for 10 different realisations of the randomly perturbed initial condition (\ref{eq:h(0) initial condition}--\ref{eq:perturbation initial condition}) on parabolic cylinder and saddle substrates rotating at $\omega = 100\,\text{rad/s}$.
While the location of the fingers depends on the individual realisation, this confirms that the observations of section~\ref{sec:non-axisymmetric substrate} are robust, that is, the fingers are consistently deflected in the $\pm x^1$ directions on the parabolic cylinder substrate and along the $x^1 = \pm x^2$ diagonals on the saddle substrate.

We note in passing that for this specific value of the angular velocity, the ensemble-averaged contact line (shown with a thick black line) spreads non-axisymmetrically: on the parabolic cylinder substrate, the contact line moves approximately 10\% faster in the $\pm x^1$ directions; on the saddle substrate, the contact line develops into a square shape.

\bibliographystyle{jfm}
\bibliography{SpinCoatingPaper}

\end{document}